\documentclass[a4paper,fleqn,usenatbib]{mnras}

\usepackage{newtxtext,newtxmath}
\usepackage{graphicx}   
\usepackage[utf8]{inputenc} 
\usepackage{multicol,rotating} 
\usepackage[section]{placeins}

\title[The Anhur/Bes regions on 67P comet]{The highly active Anhur-Bes regions in the 67P/Churyumov - Gerasimenko comet: results from OSIRIS/ROSETTA observations}
\author[S. Fornasier et al.]{
\parbox{\textwidth}{
S. Fornasier$^{1}$\thanks{E-mail: sonia.fornasier  at obspm.fr}, C. Feller$^{1}$, J.-C. Lee$^{2}$, S. Ferrari$^{3}$, M. Massironi$^{3,4}$, P. H.Hasselmann$^{1}$, J.D.P Deshapriya$^{1}$, M.A. Barucci$^{1}$, M.R. El-Maarry$^{5}$, L. Giacomini$^{4}$, S. Mottola$^{6}$, H.U., Keller$^{6,7}$, W.-H. Ip$^{8,9}$, Z.-Y. Lin$^{8}$, H. Sierks$^{10}$, C. Barbieri$^{11}$, P. L. Lamy$^{12}$, R. Rodrigo$^{13, 14}$, D. Koschny$^{15}$, H. Rickman$^{16, 17}$, J. Agarwal$^{10}$, M. A’Hearn$^{18}$, J.-L. Bertaux$^{19}$, I. Bertini$^{11}$, G. Cremonese$^{20}$, V.  Da Deppo$^{21}$, B. Davidsson$^{22}$, S.  Debei$^{23}$, M. De Cecco$^{24}$, J. Deller$^{10}$, M. Fulle$^{25}$, O. Groussin$^{26}$, P. J. Gutierrez$^{27}$, C. G\"uttler$^{10}$, M. Hofmann$^{10}$, S. F. Hviid$^{6}$, L. Jorda$^{26}$, J . Knollenberg$^{6}$, G. Kovacs$^{10}$, R. Kramm$^{10}$, E. K\"uhrt$^{6}$, M. K\"uppers$^{28}$,  M. L. Lara$^{27}$  , M. Lazzarin$^{11}$, J.J. Lopez Moreno$^{27}$, F. Marzari$^{11}$, G. Naletto$^{21,29,4}$, N. Oklay$^{6}$, M. Pajola$^{30}$, X. Shi$^{10}$, N. Thomas$^{31}$, I. Toth$^{32}$, C. Tubiana$^{10}$,  J.-B. Vincent$^{6}$ 
} 
\\~\\
Affiliations are listed at the end of the paper
}

\date{Accepted XXX. Received March 2017; in original form ZZZ}

\pubyear{2017}

\begin{document}
\label{firstpage}
\pagerange{\pageref{firstpage}--\pageref{lastpage}}

\maketitle

\begin{abstract}

The Southern hemisphere of the 67P/Churyumov-Gerasimenko comet has become visible from Rosetta only since March 2015. It was illuminated during the perihelion passage and therefore it contains the regions that experienced the strongest heating and erosion rate, thus exposing the subsurface most pristine material. In this work we investigate, thanks to the OSIRIS images, the geomorphology, the spectrophotometry and some transient events of two Southern hemisphere regions: Anhur and part of Bes.\\ 
Bes is dominated by outcropping consolidated terrain covered with fine particle deposits, while Anhur appears strongly eroded with elongated canyon-like structures, scarp retreats, different kinds of deposits, and degraded sequences of strata indicating a pervasive layering. We discovered a new 140 m long and 10 m high scarp formed in the Anhur/Bes boundary during/after the perihelion passage, close to the area where exposed CO$_2$ and H$_2$O ices were previously detected. Several jets have been observed originating from these regions, including the strong perihelion outburst, an active pit, and a faint optically thick dust plume. \\
We identify several areas with a relatively bluer slope (i.e. a lower spectral slope value) than their surroundings, indicating a surface composition enriched with some water ice. These spectrally bluer areas are observed especially in talus and gravitational accumulation deposits where freshly exposed material had fallen from nearby scarps and cliffs. The investigated regions become spectrally redder beyond 2 au outbound when the dust mantle became thicker, masking the underlying ice-rich layers.

\end{abstract}

\begin{keywords}
Comets: individual: 67P/Churyumov-Gerasimenko, Methods: data analysis, Methods:observational, Techniques: photometric 
\end{keywords}

\section{Introduction}
The Rosetta spacecraft has orbited comet 67P/Churyumov-Gerasimenko (67P) for more than 2 years providing the unique opportunity to continuously investigate its nucleus, activity, and evolution from a heliocentric distance of 4 au inbound, through its perihelion passage (1.24 au), then up to 3.5 au outbound. A large complement of scientific experiments designed to complete the most detailed study of a comet ever attempted were hosted on board Rosetta. The Optical, Spectroscopic, and Infrared Remote Imaging System (OSIRIS) instrument is the scientific camera system of the ROSETTA orbiter, and it comprises a Narrow Angle Camera (NAC) for nucleus surface and dust studies, and a Wide Angle Camera (WAC) for the wide-field coma investigations (Keller et al., 2007). 
This imaging system has enabled extensive studies at high resolution (down to 10 cm/px, and even lower during the Rosetta final descent phase) of the nucleus, showing a peculiar bilobated shape with a surface characterised by a variety of astounding morphological regions including both fragile and consolidated terrains, dusty areas, depressions, pits, boulders, talus, fractures and extensive layering  (Sierks et al., 2015, Thomas et al., 2015a, 
Vincent et al., 2015; Massironi et al., 2015). The Southern hemisphere had become visible from Rosetta only since March 2015, two months before the Southern vernal equinox, and it shows a clear morphological dichotomy compared to the Northern one, with much less variety associated with the absence of wide-scale smooth terrains (El-Maarry et al., 2015a, 2016; Giacomini et al., 2016).\\ 
The 67P nucleus has a red spectral appearance with spectral properties  similar to those of bare cometary nuclei, of primitive D-type asteroids like the Jupiter Trojans (Fornasier et al., 2007), and of the moderately red Transneptunians population (Sierks et al., 2015, Capaccioni et al., 2015). The surface is globally dominated by desiccated and organic-rich refractory materials (Capaccioni et al., 2015), and it shows some colour heterogeneities. Three kinds of terrain, ranging from the relatively bluer water ice-rich mixture to the redder ones, associated mostly with dusty regions, have been identified by visible spectrophotometry from the first resolved images acquired in July-August 2014 (Fornasier et al., 2015). Local colour and compositional heterogeneities have been identified up to the decimetre scale (Feller et al., 2016) during the closest fly-by of 14 February 2015.
 Although water is the dominant volatile observed in the coma, exposed water ice has been detected in relatively small amounts (a few percents) in several regions of the comet (De Sanctis et al., 2015, Filacchione et al., 2016a, Pommerol et al., 2015, Barucci et al. 2016, Oklay et al., 2016, 2017), and in higher amounts ($>$ 20\%) in localised areas in the Anhur, Bes, Khonsu, and Imhotep regions (Fornasier et al., 2016; Deshapriya et al., 2016; Oklay et al., 2017), and in the Aswan site (Pajola et al., 2017). Thanks to the unprecedented spatial resolution, VIRTIS and OSIRIS instruments have detected the indisputable occurrence of water frost close to the morning shadows, highlighting the diurnal cycle of water (De Sanctis et al, 2015, Fornasier et al., 2016). Water frost is strongly correlated with the speed of the receding shadows and has an extremely short lifetime of a few minutes as observed while approaching perihelion (Fornasier et al., 2016). \\
  Seasonal and diurnal colour variations of the surface of 67P's nucleus from inbound orbits to the perihelion passage have been reported by Fornasier et al. (2016). The nucleus became spectrally less red, i.e. the spectral slope decreased, as it approached perihelion, indicating that increasing activity had progressively shed the surface dust, partially showing the underlying ice-rich layer. A change in the physical properties of the outermost layer is also indicated by the evolution of phase reddening of the nucleus (i.e. the increase of spectral slope with phase angle) over time: the phase reddening coefficient decreased by a factor of two in the 2015 observations, approaching the perihelion passage,  compared to the observations acquired on August 2014, at 3.6 au inbound (Fornasier et al., 2016).\\

\begin{figure}
\includegraphics[width=\columnwidth]{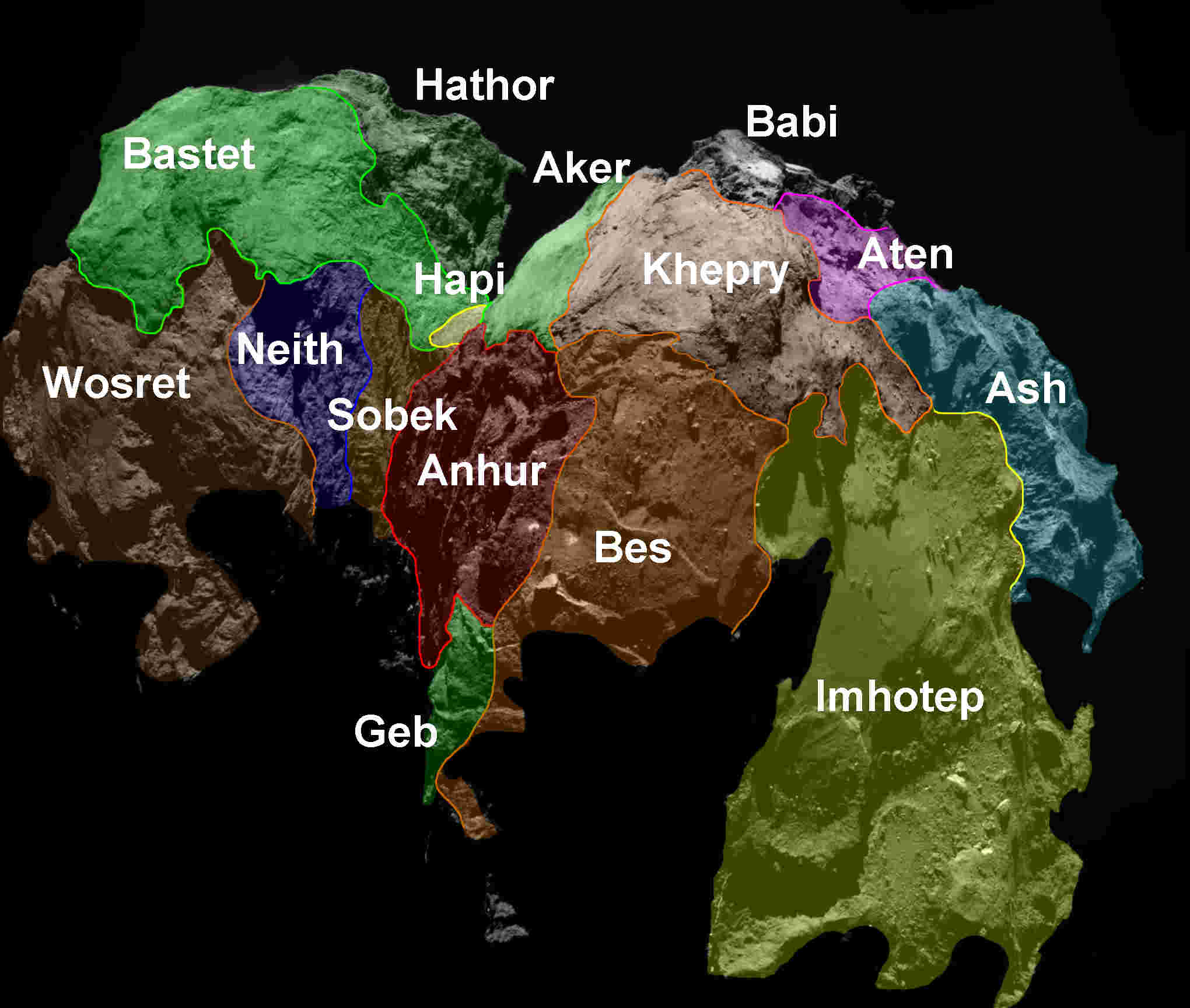}
\caption{Image from 2 May 2015 UT 07:53 showing the morphological regions visible at that time and in particular the location of the Anhur and Bes regions.}
\label{regions}
\end{figure}

In this paper we present a spectrophotometric and geomorphological analysis of two regions located in the Southern hemisphere of the 67P nucleus: Anhur and part of  Bes regions (see Fig.~\ref{regions} and El-Maarry et al. (2015a, 2016) for the description of the 67P morphological regions). These regions are more fragmented compared to other areas of the nucleus.
 They experience strong thermal effects because they are illuminated for a relatively short interval during the comet orbit, but close to the perihelion passage. These regions look particularly interesting as some large bright patches of exposed water ice have been identified pointing to local compositional heterogeneities. They are also highly active regions and sources of several jets, including the strongest outburst observed by Rosetta, which took place at the comet perihelion passage (Vincent et al., 2016a).

\section{Observations and data reduction} 

\begin{table*}
         \begin{center} 
         \caption{Observing conditions for the NAC images used in this study. The time refers to the start time of the first image in case of multiple filters observations. Filters:  F22 (649.2 nm), F23 (535.7 nm), F24 (480.7 nm), F16 (360.0 nm), F27 (701.2 nm), F28 (743.7 nm), F41 (882.1 nm), F51 (805.3 nm), F61 (931.9 nm), F71 (989.3 nm), F15 (269.3 nm) }
         \label{tab1}
        \begin{tabular}{|lclllll|} \hline
time          & Filter & $\alpha$ & r$_{\odot}$ & $\Delta$ & res & figure  \\
              &        &   ($^{\circ}$)  & (au)        &   (km)   & (m/px) &       \\ \hline
2015-05-02T07.53 & all &  61.5 & 1.732 &  125.0 &   2.4 & 1, 5, 13a \\
2015-05-02T08.53 & all &  61.5 & 1.732 &  125.0 &   2.4 & 2a,b, 5, 15 \\
2016-01-30T18.03 & F22 &  62.0 & 2.249 &  59.0  &   1.1 & 2c,b, 13d \\
2016-02-10T07.14 & F22 & 66.0  & 2.329 & 50.0   &   1.0 & 2e,f, 3 \\
2016-07-09T03.28 & F22 & 89.6 &  3.359 & 16.1   &   0.3 & 4 \\
2015-04-12T17.15 & F22, F23, F24, F16, F27, & 82.2 & 1.879 & 146.0  & 2.7 & 5, 7, 15  \\
                 & F28, F41, F51, F61, F71  &      &       &        &     & 5, 7, 15    \\
2015-04-27T18.17 & F22, F24, F41 & 73.6 & 1.763 & 134.0  & 2.5 & 5 \\
2015-04-29T17.49 & F22 & 74.4 & 1.748 & 159.7  & 3.0 & 5 \\
2015-05-02T10.42 & all &  61.5 & 1.732 &  125.0 &   2.4 & 5 \\
2015-05-07T23.35 & F22 &  60.8 & 1.688 &  137.0 &   2.7 & 5 \\
2015-03-25T02.36 & F22, F23, F24, F41 &  74.0 & 2.02 & 98.5 & 1.8 & 6 \\
2015-06-27T17.48 & F22, F23, F24, F16, F27, & 90.0 & 1.370 & 168.0  & 3.2 & 8   \\
                 & F28, F41, F51, F61, F71  &      &       &        &     & 8    \\
2016-01-27T18.36 & F22, F24, F41            & 62.0 & 2.225 & 70.0  &  1.4  &  9  \\ 
2016-02-10T08.14 & F22, F23, F24, F16, F27, &  66.0  & 2.329 & 50.0   &   1.0 &  9, 10   \\
                 & F28, F41, F51, F61, F71  &      &       &        &     & 9, 10    \\
2015-08-12T17.20 & F22, F24, F41            & 90.0 & 1.243 & 337.0 &  6.3& 11 \\
2015-08-12T17.50 & F22, F24, F41            & 90.0 & 1.243 & 337.0 &  6.3& 11 \\
2015-08-30T12.21 & all           & 70.0 & 1.261 & 405.0 &  7.6& 11, 15 \\
2016-01-27T15.20 & F22, F24, F41            & 62.0 & 2.225 & 70.0  &  1.4  &  11  \\
2016-01-27T16.27 & F22        & 62.0 & 2.225 & 70.0  &  1.4  &  11  \\
2016-01-27T17.27 & F22        & 62.0 & 2.225 & 70.0  &  1.4  &  11  \\
2015-12-07T14.03 & F41 & 89.0 &  1.833 &  102.3 & 1.9 & 12 (a,b,c) \\
2015-06-05T00.36 & all & 87.4 & 1.495 & 209.7 &  3.9  & 12d, 15 \\ 
2015-05-02T06.54 & F41 &  61.5 & 1.732 &  125.0 &   2.4 & 13b \\
2015-12-10T09.47 & F22 & 89.1 &  1.854 &  100.3 & 1.9 & 13c \\
2016-05-07T04.15 & F22, F23, F24, F27, F16, F41, F71 & 91.9 & 2.954 & 11.1 & 0.2 & 14 \\ 
2016-01-27T19.36 & all       & 62.0 & 2.225 & 70.0  &  1.4  &  15 \\
2016-06-25T11.50 & all       & 87.0 & 3.275 & 18.0  &  0.4  &  15 \\ \hline
        \end{tabular}
\end{center}
 \end{table*}

The images used for this study were reduced using the OSIRIS standard pipeline  up to level 3B, following the data reduction steps 
described K\"uppers et al (2007), and Tubiana et al. (2015).\\
Those steps include correction for bias, flat field, geometric distortion, 
and absolute flux calibration (in $W m^{-2} nm^{-1} sr^{-1}$), and finally the conversion in radiance factor (named $I/F$, where I is the observed scattered radiance, and F the incoming 
solar irradiance at the heliocentric distance of the comet), as described in Fornasier et al. (2015). \\
For the spectrophotometric analysis, the images of a given observing sequence were first coregistered taking  the F22 NAC filter (centred at 649.2nm) as reference and using a python script based on the scikit-image library (Van der Walt et al., 2014). To retrieve the radiance factor at pixel level precision for the high resolution images acquired in 2016, we also used the optical flow algorithm to improve the image coregistration (Farneback, 2001). Images were then photometrically corrected applying the Lommel-Seeliger disk law:  
\begin{equation}
D(i,e,a) = \frac{2\mu_{i}}{\mu_{e}+\mu_{i}}
\end{equation}
where $\mu_{i}$ and $\mu_{e}$ are respectively the cosine of the solar 
incidence (i) and emission (e) angles. The geometric information about the illumination and observation angles were derived using the 3D stereophotoclinometric shape model (Jorda et al., 2016) and considering all relevant geometric parameters, such as the camera distortion model, the alignment of the camera to the Rosetta spacecraft, the orientation of the spacecraft (with reconstructed orbit position and pointing) with respect to the 67P nucleus. 

RGB maps, in false colours, were generated from coregistered NAC images acquired with the filters centred at 882 nm, 649 nm, and 480 nm using the STIFF code (Bertin, 2012).

The data presented here were acquired from March 2015, when the Anhur/Bes regions became visible from Rosetta, through July 2016. Details on the observing conditions are reported in Table~\ref{tab1}.

%
                          \section{Geomorphology}

\begin{figure*}
\includegraphics[width=0.85\textwidth]{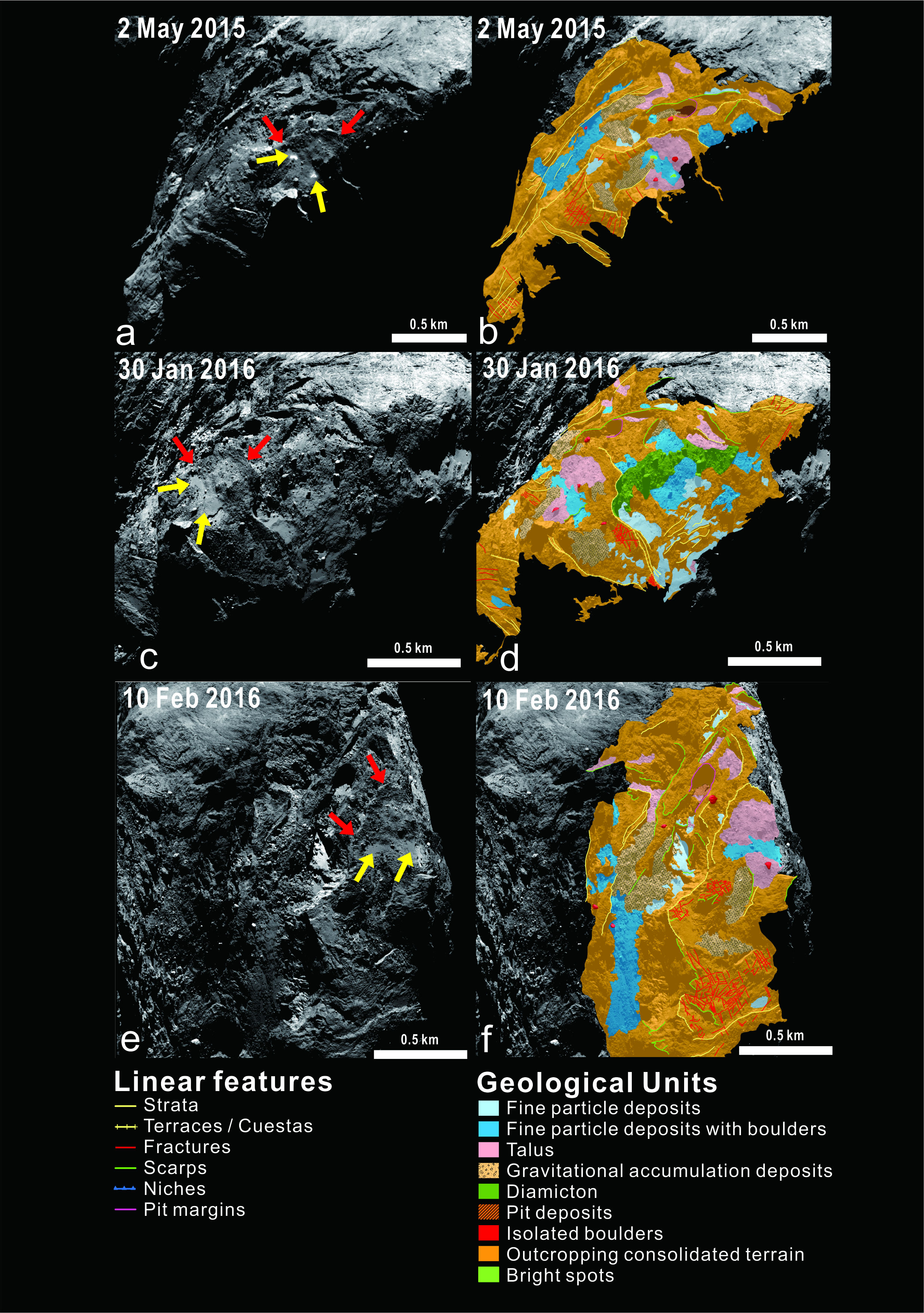}
\caption{The geomorphological maps of Anhur and Geb regions in 2 May 2015 (top panel), of Anhur and Bes regions in 30 January 2016 (medium panel), and of Anhur and Geb regions in 10 February 2016 (bottom panel). The yellow arrows indicate the flat terrace where the two bright patches were found, and the red arrows the two 150 m high scarps bounding this terrace.}
\label{maps}
\end{figure*}

\begin{figure}
\centering
\includegraphics[width=\columnwidth]{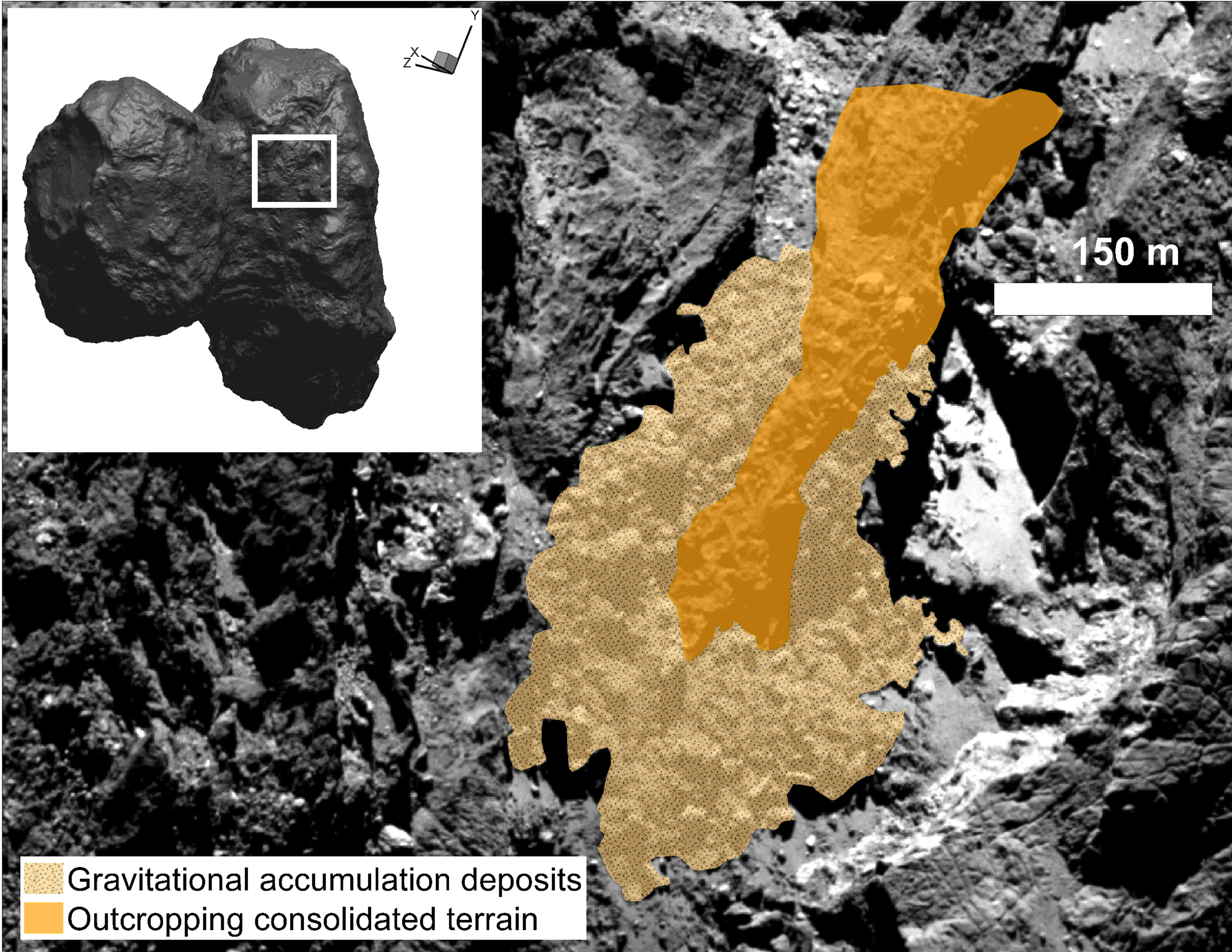}
\caption{Gravitational accumulation deposits intermixed with altered outcropping material.}
\label{maps_zoom}
\end{figure}

\begin{figure}
\centering
\includegraphics[width=\columnwidth]{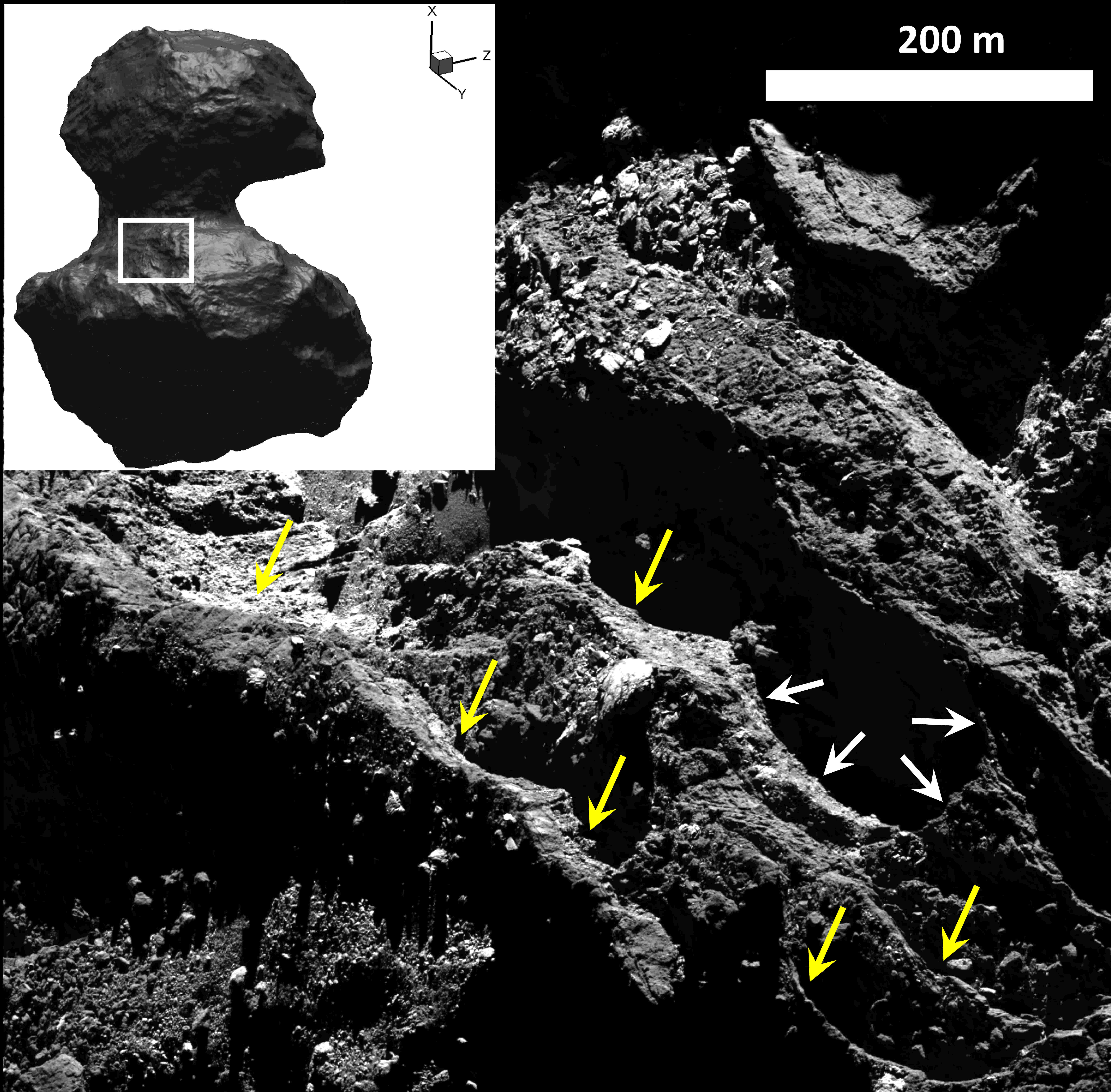}
\caption{Series of arch shaped niches (yellow arrows) and a pit rim (white arrows) complicate scarps.}
\label{maps2}
\end{figure}

We used three OSIRIS Narrow Angle Camera (NAC) images to produce the geomorphological mapping of Anhur, Bes and Geb regions (El-Maarry et al. 2016) at different viewing angles. One NAC image (Fig.~\ref{maps}, top panel) was obtained on 2 May 2015 when the comet was at the equinox, and the other two (Fig.~\ref{maps}, medium and bottom panels) were acquired during the post-perihelion phase. We crosschecked the images acquired in different periods and different viewing angles to analyse possible geomorphological changes. \\
Mapping of top and medium panels of Fig.~\ref{maps} are centered on Anhur and Bes regions, respectively, whereas the bottom panel displays both Anhur and Bes, and part of Geb. All the images show that outcropping consolidated terrains prevail and are sculpted by staircase terraces. This supports the inner stratification hypothesis (Massironi et al. 2015). By comparing the different views it is clear that, assuming an onion-like layering of the nucleus (Massironi et al. 2015), Bes is in a shallower structural level with respect to Anhur and Geb, which are dominated by an elongated canyon-like depression. The flank of the depressions displays a uniformly degraded sequence of strata. The long depression and the bounding cliffs and terraces extend from Anhur to Geb without any evident morphological interruption (Fig.~\ref{maps}, medium panel). Hence the Geb region is completely indistinguishable from Anhur at least from a structural point of view. For this reason, we will refer only to Anhur region. Both Anhur and Bes consolidated terrains are crossed in places by long fracture systems superimposed on the m-scale polygonal ones already described by El-Maarry et al. (2015b, 2016) on other regions of the comet nucleus.    

The strata and cuesta margins in Anhur region have apparently been dissected by several scarps with various kinds of deposits at their feet. In particular we have distinguished talus and more heterogeneous gravitational accumulation deposits. They are both interpreted as made up of material dislodged from adjacent cliffs due to concurrent sublimation and gravitational processes (Pajola et al. 2015, La Forgia et al. 2015, Vincent et al. 2016b). We have also distinguished diamictons that are characterised by a highly heterogeneous texture with blocks and fine materials as the gravitational accumulations, but unlike the latter, diamictons are located far from the cliffs. The diamictons can be leftovers from erosive scarp retreats (Pajola et al. 2016) or the result of in-situ degradation of the underlining consolidated material (Fig.~\ref{maps}). In some cases even gravitational accumulation deposits at the scarp feet can be intermixed with altered outcropping material with evident breaks and voids (Fig.~\ref{maps_zoom}). This might suggest an in situ degradation more effective than previously supposed.\\
Fine materials, occasionally associated with some isolated boulders (the major ones, larger than 18 m, even resolved within the maps) are scattered throughout the analysed regions and preferentially deposited on flat terrains. The thickness and extension of such deposits could be very variable and in some cases they are thin enough to let the underlined consolidated material be easily inferred. The fine material deposits are likely formed by airfalls (Thomas et al. 2015; Lai et al., 2016). Boulders can be erratic, i.e. transported from steep slopes onto flat terrains through out-gassing and gravitational coupling (e.g. El-Maarry et al. 2015, Lee et al., 2016) or simply be the result of in-situ degradation (Lee et al., 2016).\\
Scarps are often complicated by arch-shaped niches (Fig.~\ref{maps2}). They could be simply crown areas of gravitational falls or remnants of former pits.



\section{Composition and colour changes}

\subsection{Large exposed water ice-rich patches}

\begin{figure*}
\centering
\includegraphics[width=1.0\textwidth]{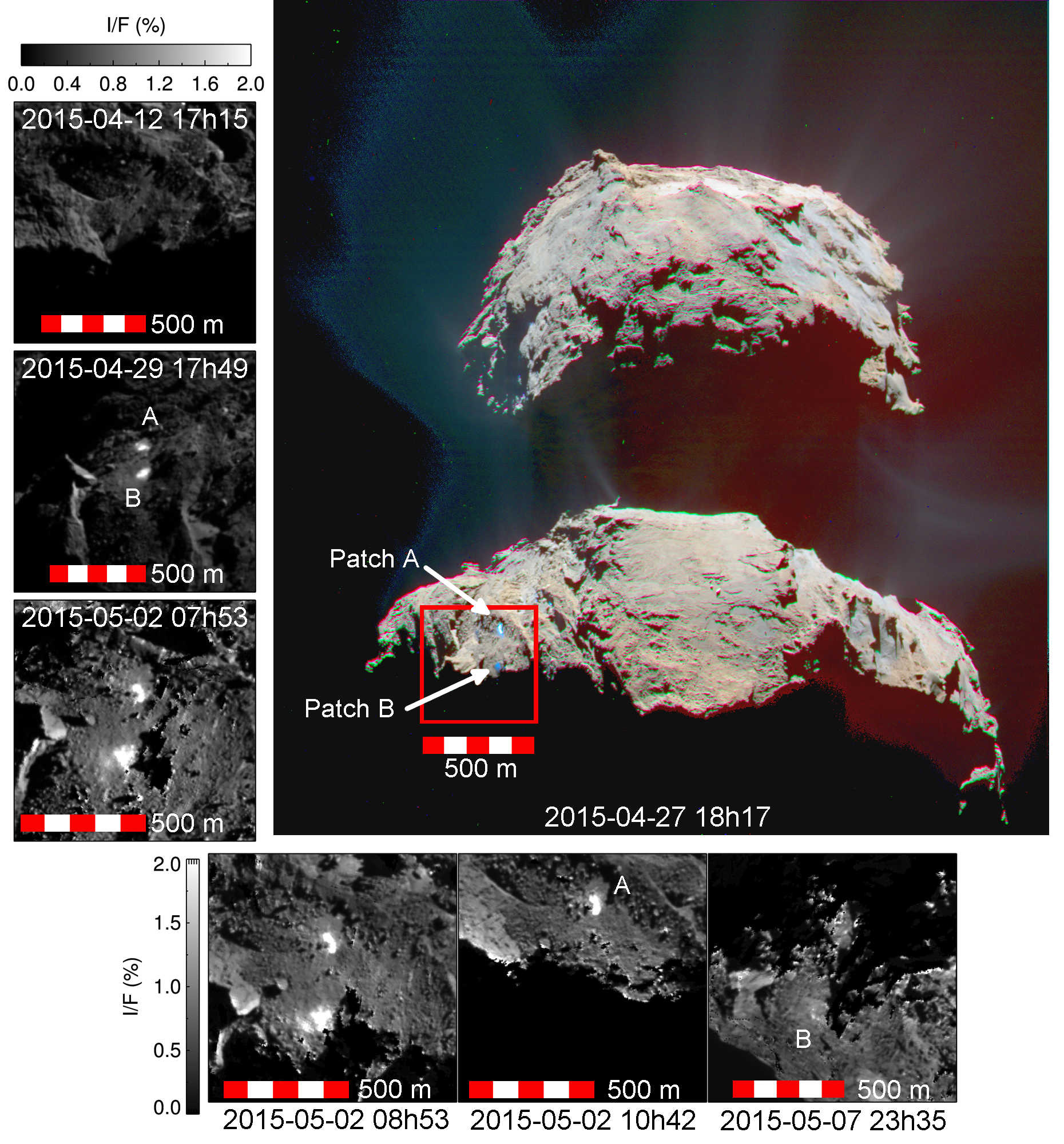}
\caption{Composite RGB image (using filters centred at 882, 649, and 480 nm) showing the appearance of the bright patches in the Anhur/Bes regions on 27 April 2015, UT18h17 and the surface evolution in the 12 April-7 May 2015 time frame.}
\label{patches27April}
\end{figure*}

Two bright patches of about 1500 m$^2$ each were revealed in the Anhur and Bes regions (Fig.~\ref{patches27April}), near their boundary (Fornasier et al., 2016). These areas are 4-6 times brighter than the comet dark terrain and they have a spectral behaviour completely different from that of the nucleus dark terrain. Their spectrophotometry is flat and consistent with a water ice abundance of about 20\% (patch B, Fig.~\ref{patches27April}) and 32\% (patch A), according to the linear mixing model presented in Fornasier et al. (2016). This water-ice abundance is considerably higher than the few percent measured locally by the VIRTIS instrument (Filacchione et al., 2016a; De Sanctis et al., 2015, Barucci et al., 2016). The bright patches appeared for the first time on images acquired on 27 April 2015 (Fig.~\ref{patches27April}), and they were still clearly visible on 29 April and 2 May 2015 images. Patch A in Fig.~\ref{patches27April} is located at longitude  [65.8$^{\circ}$, 67.8$^{\circ}$] and latitude  [-54.7$^{\circ}$, -53.9$^{\circ}$] and it is about 25\% brighter and two times bigger (about 1200 m$^2$) than patch B on 27 April images. Patch B  is located at longitude [76.3$^{\circ}$,76.6$^{\circ}$] and latitude [-54.3$^{\circ}$,-54.0$^{\circ}$], and progressively increases in size from 27 April to 2 May images, indicating a later exposure of water ice than patch A, whose size decreased from 29 April to 2 May images (see Table S2 in Fornasier et al. (2016) for the size evolution of the patches). On images acquired on 7 May 2015, only patch B was still visible while patch A had fully sublimated. In the 27 April to 2 May images, patch B has a fuzzy appearance indicating an ongoing sublimation.  From the observed water ice abundance of about 25\%, and assuming a dust/ice bulk density ratio of 2, and a similar porosity for dust and ice material, we derive a dust/ice mass ratio of 6, consistent with the value found by Fulle et al. (2016) for the comet's inbound orbit.

Subsequent images covering the region of interest were acquired only on 5 June 2015, revealing that the water ice patches had completely sublimated and that the spectrophotometry of those regions previously covered by the bright patches was totally indistinguishable from that of the surrounding comet dark terrain (see Fig. 2 in Fornasier et al. (2016) for the spectrophotometric evolution of these surfaces). 

\begin{table*}
         \begin{center} 
         \caption{Thermal modelling for the region containing the 2 bright ice rich patches (see Fig.~\ref{patches27April}). $A_B$ is the bond albedo, lat$_{s}$ the subsolar latitude.} 
         \label{thermalmodeling}
        \begin{tabular}{|lllllll|} \hline
day                & r$_{\odot}$  & lat$_{s}$ & A$_B$ & prod. rate  & erosion  &  T$_{max}$  \\
                     &     (au)         &  ($^{\circ}$)      &            & (kg/m$^2$/d)    & (mm/d)      & (K) \\ \hline
12/04/2015    & 1.8791         &  9.9         & 0.05     & 1.43            & 2.68     & 193.50            \\
12/04/2015    & 1.8791         &  9.9         & 0.70     & 0.14            & 0.26     & 179.48            \\                     
12/04/2015    & 1.8791         &  9.9         & 0.01     & 1.12            & 2.11     & 208.83            \\
24/04/2015    & 1.7880         &  5.9         & 0.05     & 1.90            & 3.56      & 194.70            \\
24/04/2015    & 1.7880         &  5.9         & 0.70     & 0.23            & 0.44     & 181.67          \\
24/04/2015    & 1.7880         &  5.9         & 0.01     & 1.53             & 2.87     & 212.01          \\
07/05/2015    & 1.6915         &  1.1         & 0.05     & 2.55            & 4.79      & 195.59          \\
07/05/2015    & 1.6915         &  1.1         & 0.70     & 0.38            & 0.71      & 184.14           \\
07/05/2015    & 1.6915         &  1.1         & 0.01     & 2.10            & 3.95     & 215.64           \\ \hline
        \end{tabular}
\end{center}
 \end{table*}

\begin{figure*}
\centering
\includegraphics[width=1.0\textwidth]{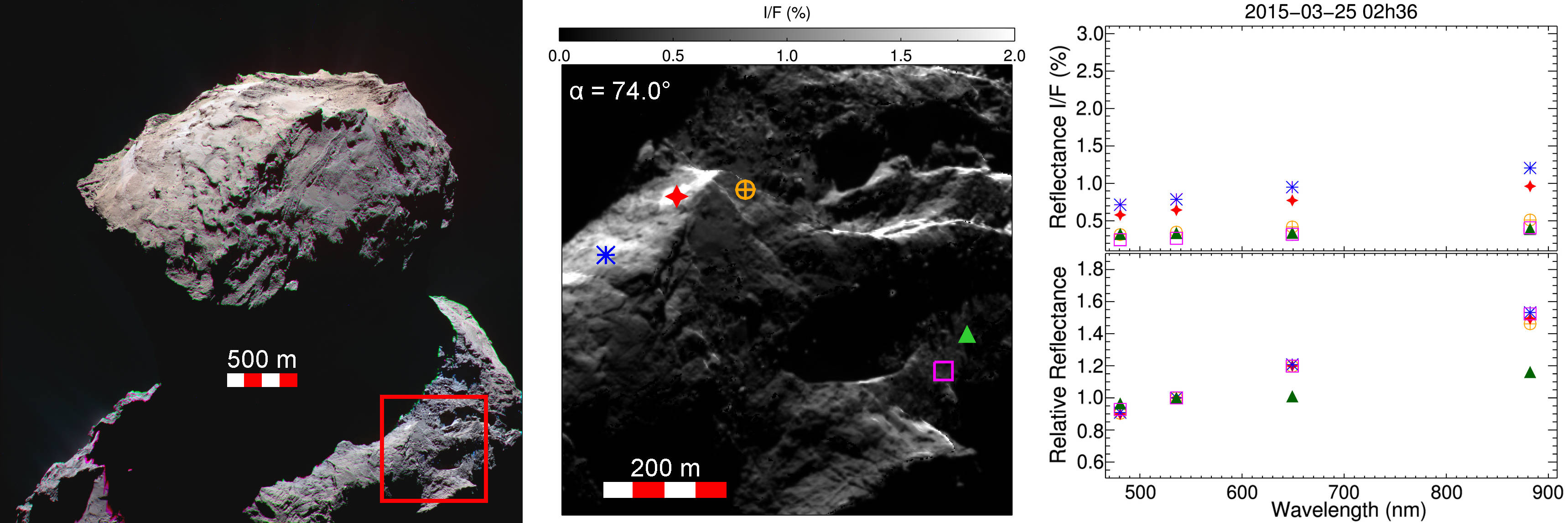}
\caption{Composite RGB map (using filters centred at 882, 649, and 480 nm) from 25 March 2015 02h36 images (left), and spectrophotometry of 5 selected regions of interest. The reflectance is given at phase angle 74$^{\circ}$. The Sun is toward the top.}
\label{spec_25march}
\end{figure*}

The bright water ice-rich regions were located on a flat terrace bounded by two 150 m high scarps (Fig.~\ref{maps}). The overall structure suggests a strata plane covered by different deposits. In particular the bright patches were formed on a smooth terrain made up of a thin seam of fine deposits covering the consolidated material, which, indeed, outcrops nearby and displays a polygonal network of fractures. Chaotic and heterogeneous terrains classified as gravitational accumulation or diamictons surround the fine materials. Diamictons in this location are more probably the result of in situ differential erosion on the consolidated material since on high-resolution images these deposits are strictly intermixed with outcropping materials. 

In order to estimate the amount of water ice present in the bright patches we performed thermal modelling
of the Anhur/Bes region. The optical and thermal properties of the surface are strongly affected by the
type of mixing of the refractory material with the ice. Because the type of
material mixing on a sub-resolution scale 
has not yet been established for this comet, we studied two end cases. For the first one we assumed that 
the surface consists of an intimate mixture of dust and ice, with the two components mixed at sub-mm 
scales and assumed to be in thermal equilibrium  (model A of Keller et al. 2015). For this case we used 
a Bond albedo of 0.05, i.e. a value 5 times higher than the one measured at 535 nm for
67P by Fornasier et al. (2015), in order to match the higher albedo observed for the ice-rich patches.

For the second end case we considered a surface consisting of pure ice and refractory material well separated 
on small scales -- but still spatially unresolvable by OSIRIS -- and non thermally interacting. For this case the 
brightness observed by OSIRIS corresponds to the linear combination by surface of the two individual 
components. In this case we used thermal model A with a Bond albedo for pure water-ice snow of 0.7 (Delsemme, 1985).

For the purpose of comparison, we also modeled the behavior of the dark background material
with model B (Keller et al. 2015), which describes the comet surface as a two-layer system,
in which a subsurface dust-ice mixture is overlain by a desiccated dust coating with a Bond albedo of 0.01.

The results shown in Table~\ref{thermalmodeling}, predict erosion rates ranging from about 0.25 mm/day to 5 mm/day for the facets on the shape model representative of the patches location.
The higher erosion rate corresponds to the intimate mixing model, in which, due to the lower albedo, more
input energy is available for sublimation. Conversely, the lower erosion rate corresponds to the end case in
which pure ice is mixed with refractory material at geographical, unresolved scales.
The erosion is computed for pure porous ice with a density of 533 kg/m$^3$, corresponding to the bulk density measured for 67P (Paetzold et al. 2015; Jorda et al., 2016; Preusker et al., 2015). 
Assuming a dust to water ice ratio of 4:1 (Rotundi et al., 2015), the total erosion
would be in the range 1.3 to 24 mm/day, with an estimated ice loss rate ranging from about 0.14 to 2.5 kg~d$^{-1}$~m$^{-2}$. Considering 
that the observed permanence time of the ice patches was of the order of 10 days,
the thickness of the porous dust/water ice layer is estimated to be between 13 and 24 cm for the intimate mixture case, 
while for an areal mixture of dust and ice, the original thickness would be between 13 and 36 mm.
\begin{figure*}\centering
\includegraphics[width=1.0\textwidth]{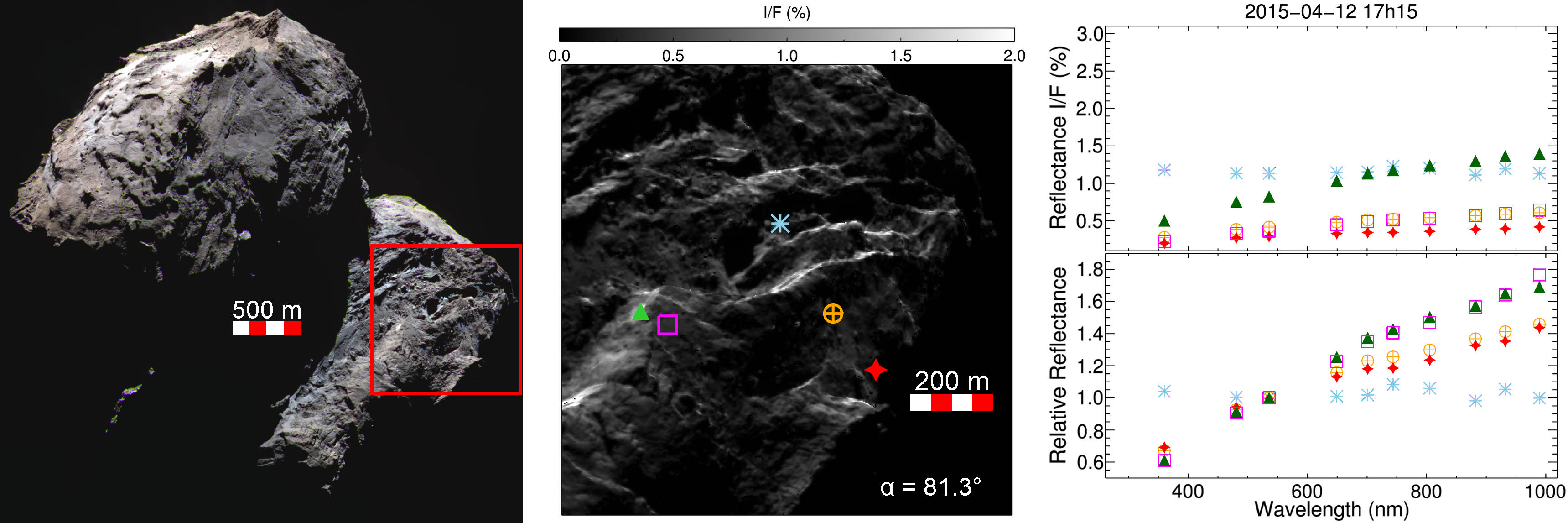}
\caption{Composite RGB map (using filters centred at 882, 649, and 480 nm) from 12 April 2015 17h15 images (left), and spectrophotometry of
 5 selected regions of interest. The reflectance is given at phase angle 81$^{\circ}$. The Sun is toward the top.}
\label{april2015}
\end{figure*}
\begin{figure*}
\centering
\includegraphics[width=1.0\textwidth]{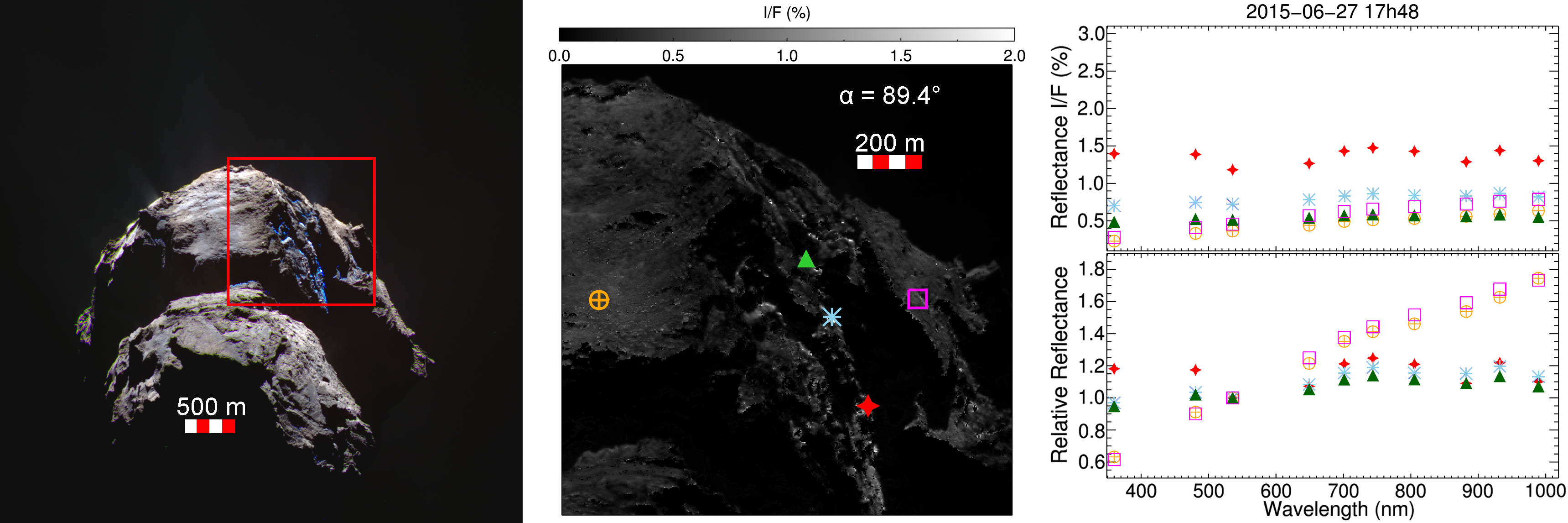}
\caption{Composite RGB map (using filters centred at 882, 649, and 480 nm) from 27 June 2015 17h48 images (left), and spectrophotometry of 
5 selected regions of interest. The reflectance is given at phase angle 89$^{\circ}$. The Sun is toward the top.}
\label{june2015}
\end{figure*}
These ice rich patches indicate a variation of the water ice content in the uppermost layers pointing to local compositional  heterogeneities of the 67P nucleus. Fornasier et al. (2016) proposed that these compositional heterogeneities are related to the re-condensation of volatiles and sintering of the subsurface material during previous perihelion passages. The uniqueness of the observed bright patches is probably related to a delicate balance between activity and dust removal leading to an icy surface for a limited duration. 

Notably, the presence of a small amount of CO$_2$ ice was detected by the VIRTIS spectrometer in a 80 m by 60 m  area corresponding to the location of patch A on 21-23 March 2015 (at longitude 66$^{\circ}$ and latitude 54.6$^{\circ}$), ice that had sublimated in less than 3 weeks (Filacchione et al., 2016b), well before the detection of the bright patches with OSIRIS. Filacchione et al. (2016) estimated the thickness of the CO$_2$ ice layer to be $<$ 5.6 cm, and propose that it was formed by the freezing of CO$_2$ at low temperature after the previous perihelion passage. In the OSIRIS archive there are no images covering the Anhur region simultaneously to the VIRTIS CO$_2$ ice detection. That region is visible in images acquired on 25 March 2015 (Fig.~\ref{spec_25march}, green triangle), located close to shadows and it has a spectrophotometric behaviour in the 480-881 nm which is quite different from the other selected regions, with a much lower spectral slope value consistent with the presence of some ice. However, this region is not brighter than its surroundings, indicating that the exposed fresh layer of CO$_2$ ice had partially sublimated and that the residual ice must be well mixed with the surface dust.

In Figure~\ref{spec_25march} a bright structure is visible, however its spectrophotometry  (red star symbol) indicates a red spectral behaviour indistinguishable from that of the comet dark terrain. Other bright and spectrally reddish regions have been observed at higher spatial resolution by OSIRIS (Feller et al., 2016). These features may be associated with brighter minerals and/or with different structure/grain size, and not with an enrichment in water ice of the surface composition.

\subsection{Spectrophotometry of smaller bright spots}

\begin{figure*}\centering
\includegraphics[width=1.0\textwidth]{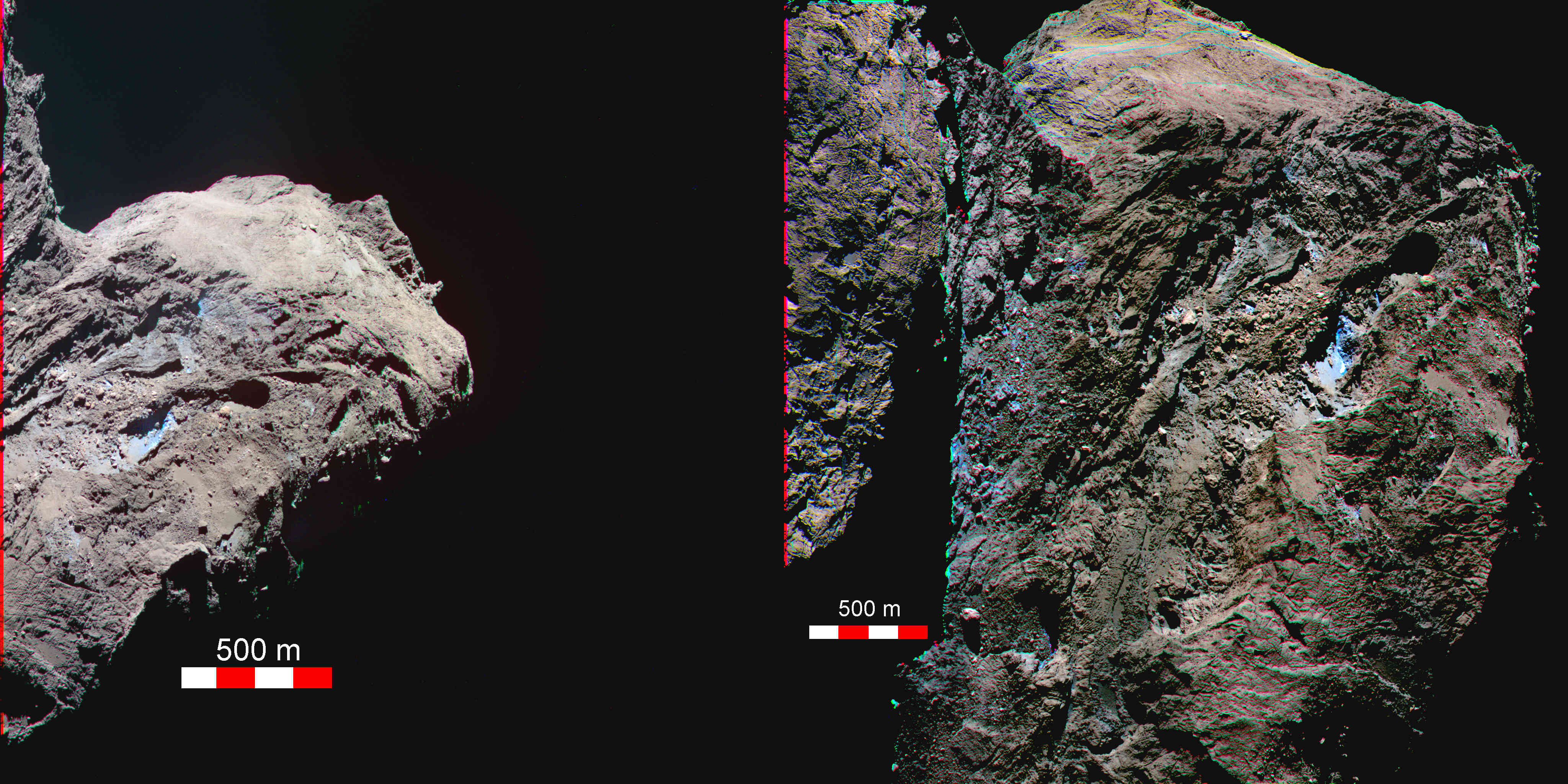}
\caption{Composite RGB map (using filters centred at 882, 649, and 480 nm) from  27 January 2016 UT 18h36 (left) and 10 February 2016 UT 08h14 (right). The Sun is toward the top. }
\label{RGB2016}
\end{figure*}

\begin{figure*}\centering
\includegraphics[width=1.0\textwidth]{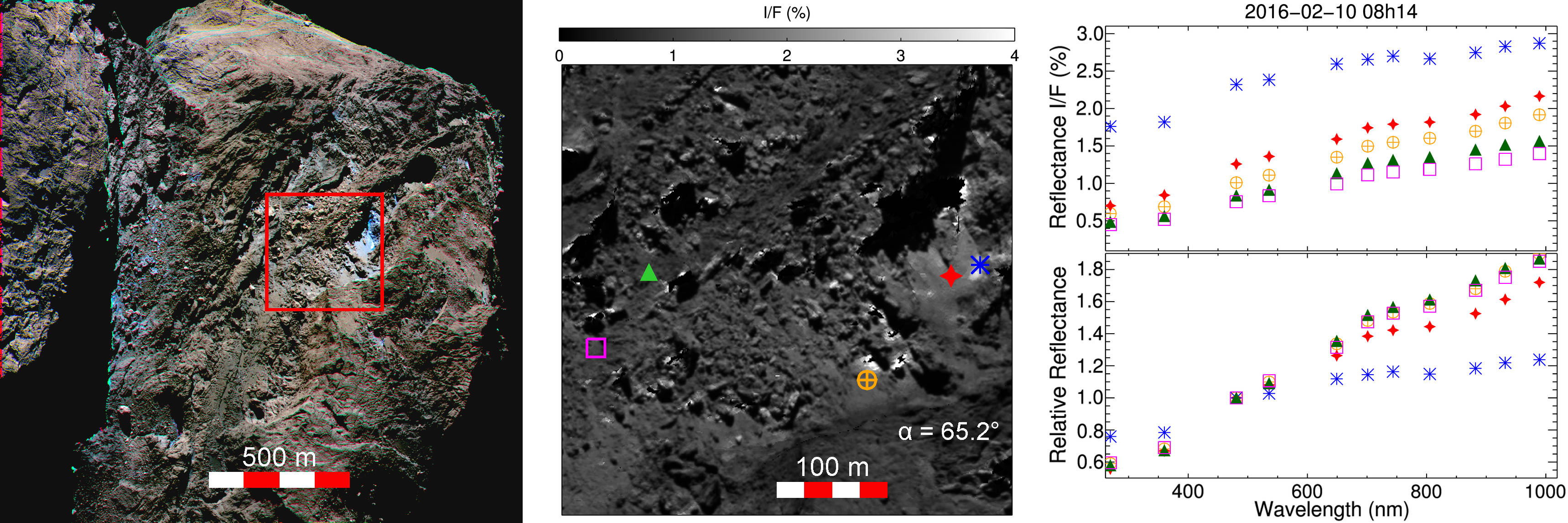}
\caption{Composite RGB map (using filters centred at 882, 649, and 480 nm) from 10 February 2016 08h14 images (left), and spectrophotometry of 5 selected regions of interest. The reflectance is given at a phase angle of 65$^{\circ}$. The Sun is toward the top. The bright structures close to shadows near the scarps should be discarded as they results from an imperfect correction of the illumination conditions.}
\label{feb2016}
\end{figure*}

Besides the two large bright patches, several smaller and localised areas enriched in water ice have been identified in Anhur, close to shadows and/or in the pit deposits, and characterised by a relatively blue spectrophotometry and an increase in the reflectance (Figs~\ref{april2015}, ~\ref{june2015}, ~\ref{RGB2016}, and  ~\ref{feb2016}). Several studies with the VIRTIS spectrometer have proven that regions having a flat or less steep spectral behaviour than the surrounding  in the visible
range have a higher abundance of water ice mixed with the refractories (Filacchione et al., 2016a; Barucci et al., 2016). For comets 9P/Tempel 1 and 103P/Hartley 2, relatively bluer colours were also associated with a higher abundance of water ice (Sunshine et al., 2006, 2012).

In Figure~\ref{april2015} the light blue symbol indicates a bright feature whose spectrophotometry is flat with a reflectance 5-6 times higher in the 360-480 nm range than the surrounding terrains, similarly to what has been observed in other ice-rich features. The bright but spectrally red feature observed on 25 March images is still present (green triangle in Fig.~\ref{april2015}) and shows a spectrum similar to that of surrounding dark terrains (magenta square). The areas where the two large bright patches A and B have later appeared (orange circle and red star) look spectrally bluer than the surroundings, indicating a progressive local enrichment in the surface water-ice abundance from the subsurface layers. \\
\begin{figure*}
\centering
\includegraphics[width=0.9\textwidth]{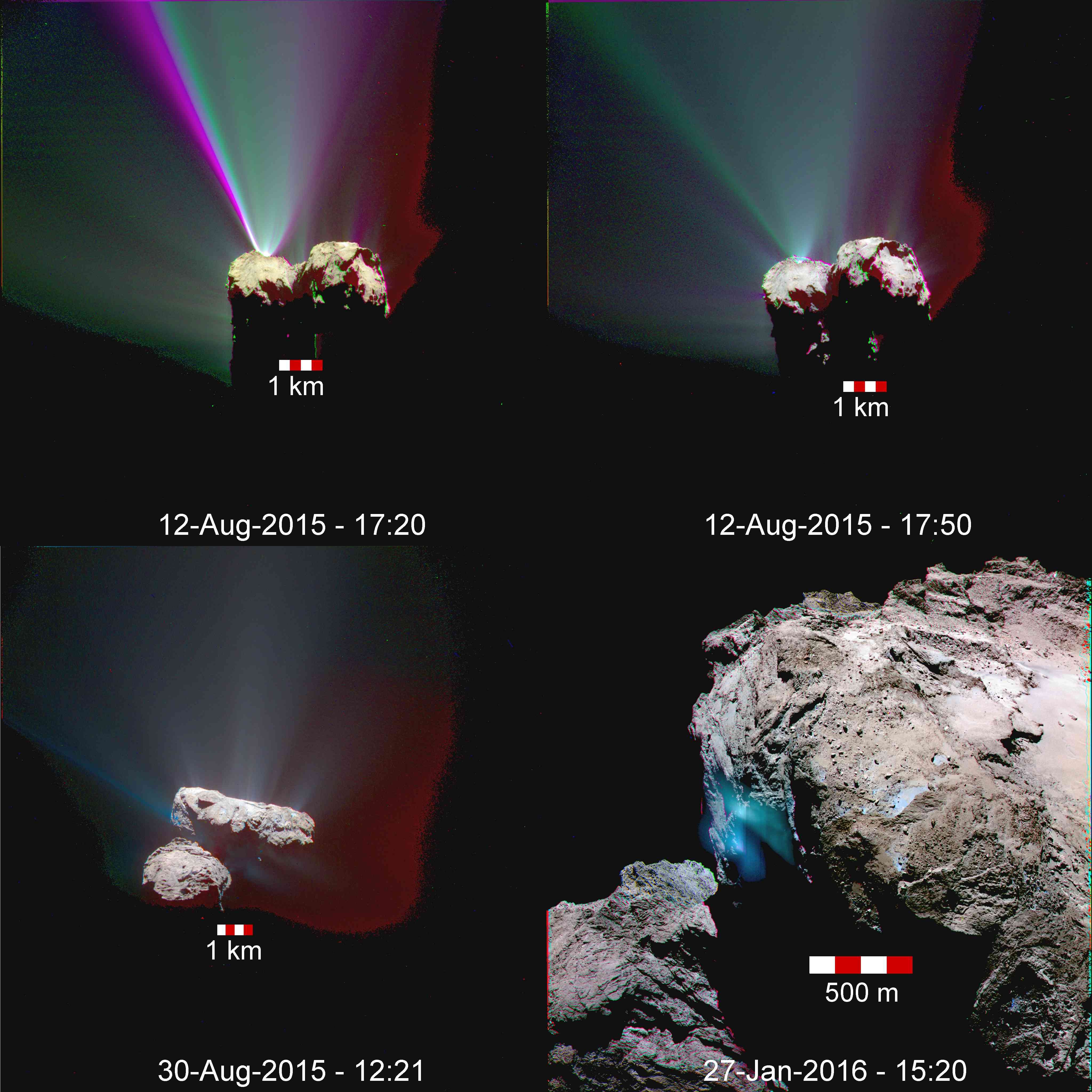}
\includegraphics[width=0.9\textwidth]{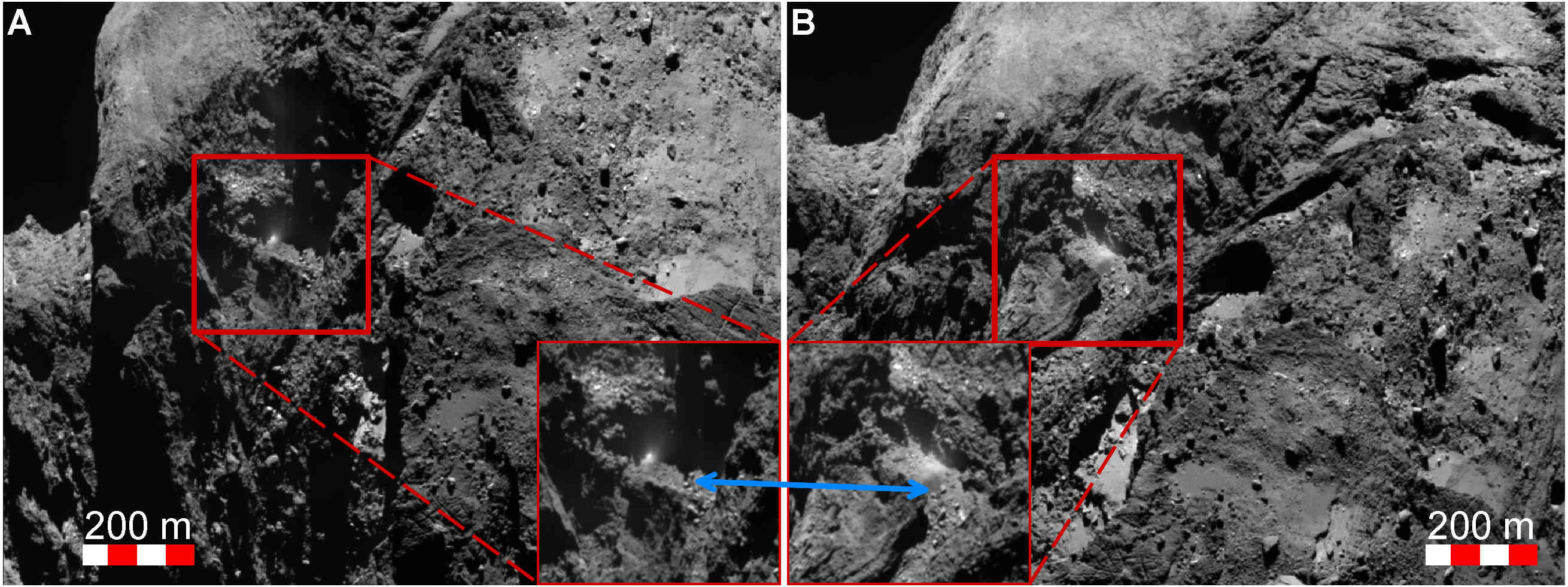}
\caption{RGB maps showing jets and activity originated from the Anhur region. On the Top: the 12 August 2015 outburst 17:20 (left), and  17:50 right;  middle panel: 30 August 2015 (UT 12:21) jets departing from Anhur region (left); jets seen on 27 January 2016 (UT15:20) images; bottom panel: images from 27 January 2016, showing localised jets (left: UT16:27; right: UT17:27, the shadow of the plume is visible at its bottom ).}
\label{outburst}
\end{figure*}
A nice view of water ice enriched regions observed close to the Anhur morning shadows is given in Fig.~\ref{june2015} (left side). The spectrophotometry in these regions indicates a flat spectral behaviour, consistent with the presence of  water ice, and, for the brightest region represented by the red star symbol, and enhancement of the flux in the UV region possibly due to the presence of some water frost. For this region (in particular for the area represented by the green triangle), visible and NIR spectra were obtained 3 months earlier with the VIRTIS spectrometer and presented in Barucci et al. (2016), indicating the presence of some exposed water ice although only a small amount (less than 1\%). \\
\begin{table*}
\caption{Locations of the jets seen in Fig~\ref{outburst}. For the 27 January 2016 UT 15:20 image the approximate positions of the two brightest jets are given. Other fainter jets are visible but their source region on the nucleus is not easy to be clearly identified.}
\label{jetpos}
\begin{tabular}{c|c|c|c|c|c|c|c}\hline
image  & $r_{\odot}$ (au) & $\Delta$ (km) & phase ($^{\circ}$)& res. (m/px) & lon ($^{\circ}$)& lat($^{\circ}$) \\ \hline
2015-08-30T12h21 & 1.26 & 405 & 70.0 & 7.6 & 49.1   &  -31.2 \\
2015-08-30T12h21 & 1.26 & 405 & 70.0 & 7.6 & 43.5   &  -31.3 \\
2015-08-12T17h20 & 1.24 & 337 & 90.0 & 6.3 & 58.9    &  -40   \\
2016-01-27T15h20 & 2.23 & 70 & 62.0 & 1.4 & $\sim$52    &  $\sim$ -29   \\
2016-01-27T15h20 & 2.23 & 70 & 62.0 & 1.4 & $\sim$59    &  $\sim$ -29   \\
2016-01-27T16h27 & 2.23 & 70 & 62.0 & 1.4 & 53.6    &  -34.6   \\
2016-01-27T17h27 & 2.23 & 70 & 62.0 & 1.4 & 56.4    &  -35.7   \\ \hline
\end{tabular}
\end{table*}

\begin{figure*}
\includegraphics[width=1.0\textwidth]{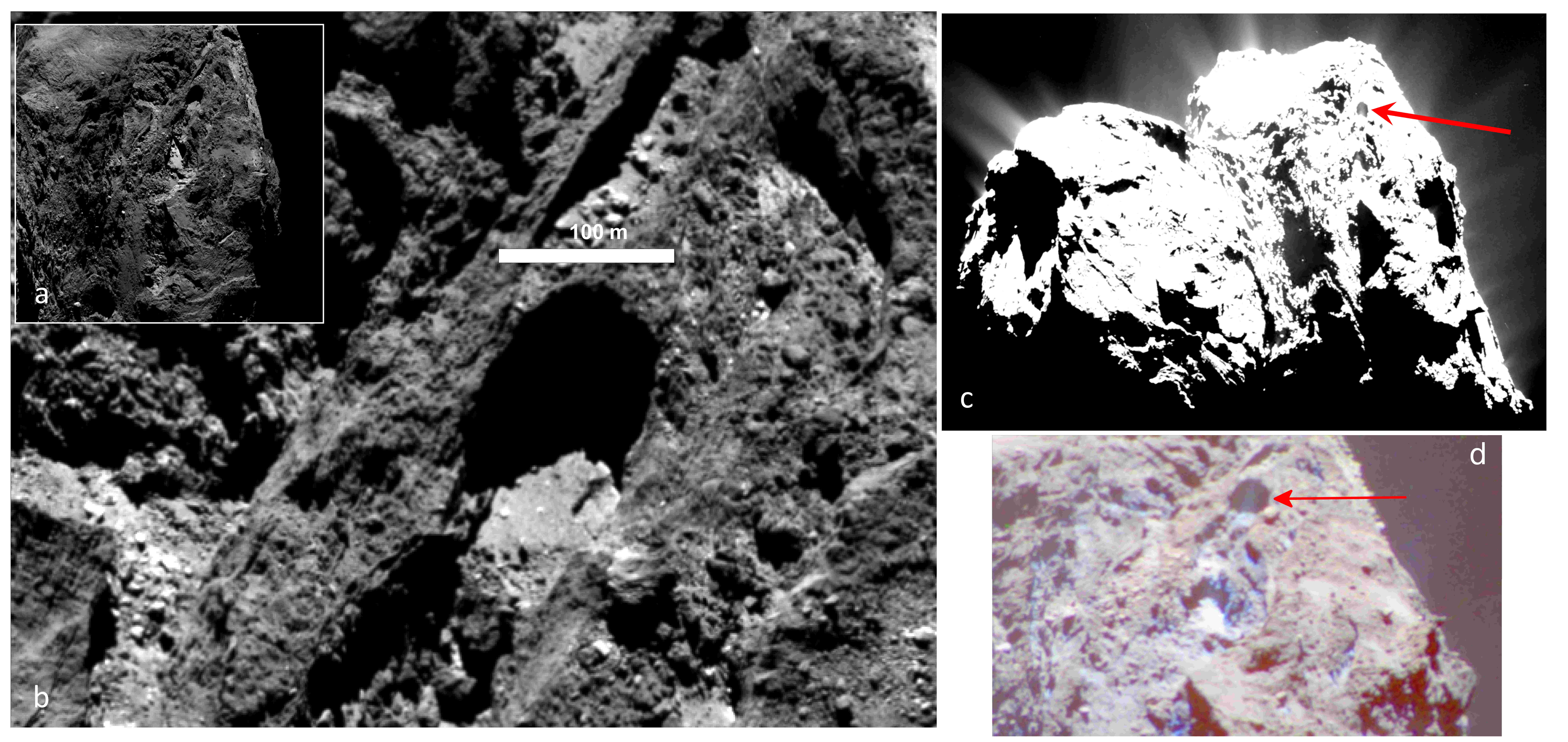}
\caption{Pit on Anhur. The pit is about 100 metres across, and we can find smooth deposits inside the pit. (c) and (d) show the activity of this pit from images acquired on 7 December 2015 and 5 June 2015, respectively. (b) is the zoomed image of (a).}
\label{activepit}
\end{figure*}
Looking at the spectrophotometry in Fig.~\ref{june2015}, we also notice an increase of the flux in the 700-750 nm region and at 930 nm, which has already been previously observed in different regions of the comet, and attributed to cometary emissions of $\rm H_2O^+$ and $\rm NH_2$ (Fornasier et al., 2015). Notably, beside the cometary emission feature, no diagnostic absorption bands are observed in any of the spectrophotometric data here presented. In particular we do not detect any absorption band associated with hydrated silicates like the one centred at 0.7 $\mu$m seen on the spectra of several hydrated asteroids (Fornasier et al., 2014), and of some transneptunian bodies (Fornasier et al., 2004, 2009, de Bergh et al., 2004). Thus the apparent absence of phyllosilicates in the Northern hemisphere pointed out by Davidsson et al. (2016) is confirmed also in the Southern hemisphere regions.

More examples of exposed water ice are shown in Figs.~\ref{RGB2016}, and ~\ref{feb2016} in the deposits of fine material close to a scarp from 10 February 2016 observations. Looking at the spectrophotometry, the fine deposits close to the scarp are relatively bluer than the other smooth regions investigated. In particular the brightest area indicated by the blue asterisk symbol is also the least steep one in terms of spectral slope, clearly indicating the exposure of water ice.

\section{Jets and activity}

Anhur is one of the most active regions on the cometary surface, and in particular it is the source of the perihelion outburst (Vincent et al., 2016a), the most intense activity event on the 67P comet recordered by Rosetta observations. Figure~\ref{outburst} shows this spectacular event that we observed on 12 August 2015 at UT 17:20, with intensity fading out in half an hour. Other jets departing from Anhur region were observed on 30 August 2015, shortly after the perihelion passage, but also several months later, on January 2016 at a heliocentric distance of 2.23 au outbound (Fig.~\ref{outburst}). Table~\ref{jetpos} gives the position, in longitude and latitude, of the region on the nucleus where the jets originated. Notably on 27 January 2016 images, diffuse jets are seen departing from several sources inside the Anhur region (UT 15:20). The activity progressively decreased. In the following couple of hours two localised jets, close but originating from different positions, are observed near the Anhur and Aker boundary. In particular, the one seen on the image acquired at 17h27 (Fig.~\ref{outburst}B), produced a dust plume casting a shadow on the surface. The plume was about 1.7 times brighter than the surrounding regions. Comparing the reflectance of the terrain before and during the shadow cast by the plume, we measure a decrease of 35\% of the reflectance in the shadow. This implies that the plume is optically thick, with an estimated optical depth of $\sim$ 0.43.
\begin{figure*}
\includegraphics[width=1.0\textwidth]{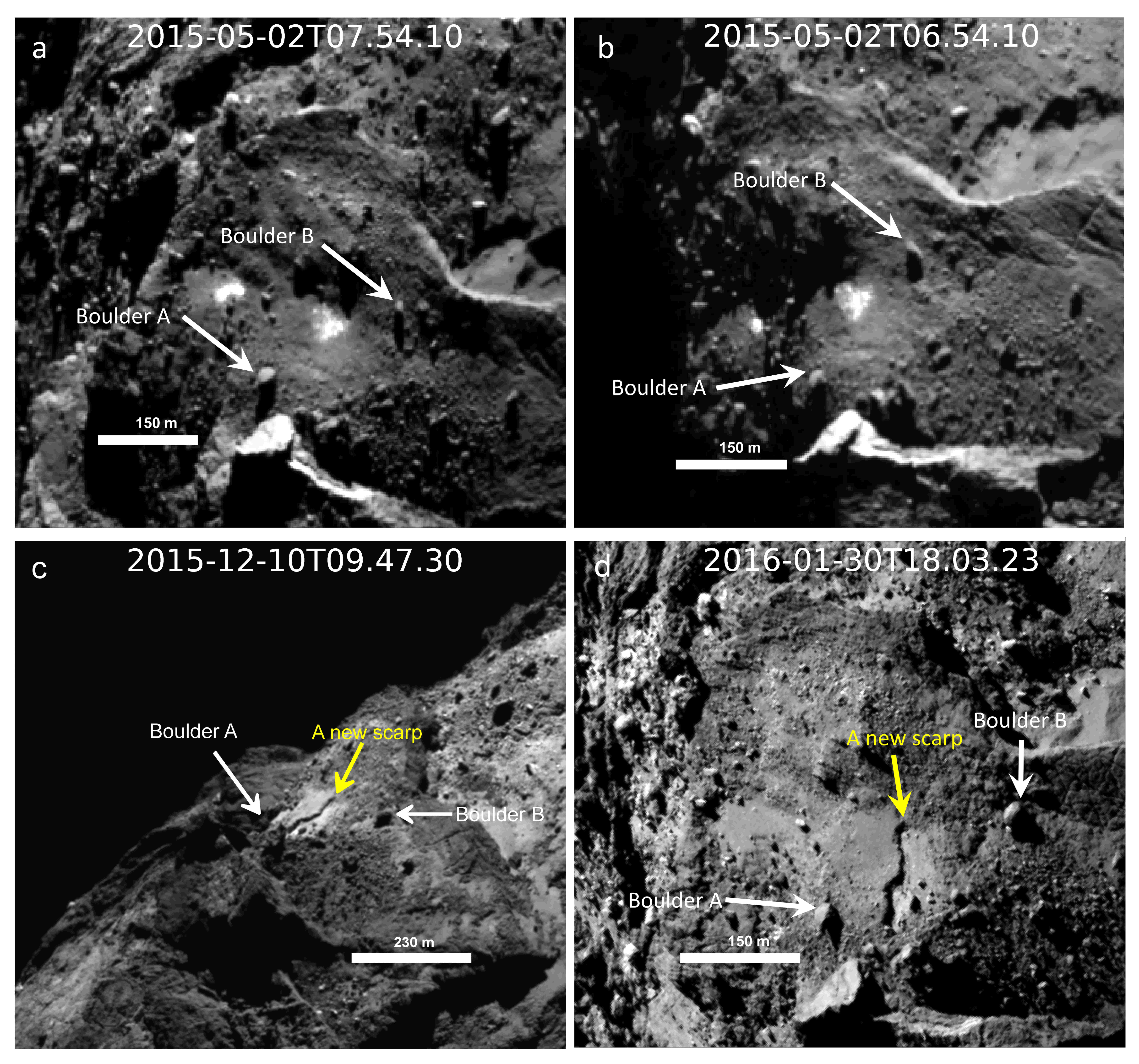}
\caption{Surface changes near the bright spots. (a) and (b) are the images taken on 2 May 2015, (c) the image acquired on 10 December 2015, where the new scarp became visible, and (d) is an image acquired on 30 January 2016, showing the scarp at a higher spatial resolution. The white arrows point out the boulders A and B that are the same boulders and the yellow arrow indicates the newly formed scarp.}
\label{scarp}
\end{figure*}

The Southern hemisphere of the comet has fewer pits compared to the Northern one. The Seth region in the Northern hemisphere is the most enriched of this kind of morphological structure, and some of them have been shown to be active (Vincent et al., 2015). The Anhur region shows the only pit proven to be active in the Southern hemisphere. It is a pit, 100 m across, whose floor is covered by very smooth deposits which are spectrally bluer than the nearby areas. Figure~\ref{activepit} shows this active pit in images acquired on 5 June 2015 (panel c), and on 7 December 2015 (panel d).
 The bluer colours of the pit floor may be produced by fresh material resulting from possible lateral outflow, as observed by Vincent et al. (2015) in some Northern hemisphere pits, and/or to floor contamination due to fall-back ejecta (dusty-ice grains) which were not able to escape far enough from the pit source.

\subsection{Morphological changes: a new scarp}

 \begin{figure*}
\centering
\includegraphics[width=1.0\textwidth]{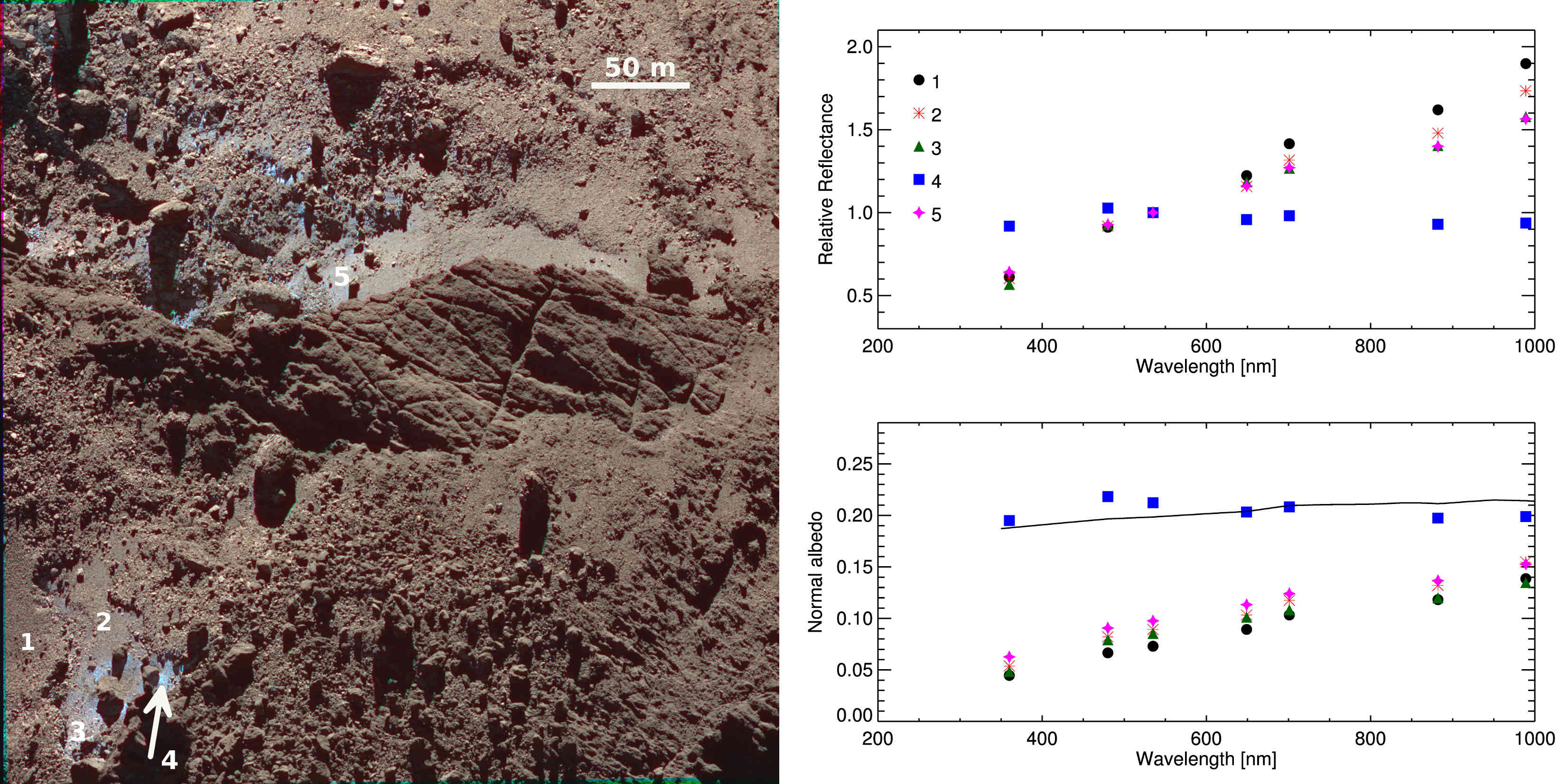}
\caption{Left panel: RBG colour map of a part of the Bes region from images acquired on 7 May 2016, UT 04:15, showing the new scarp, which formed between August and December 2015. Right panel: relative spectrophotometry of the 5 ROIs. The continuous line is an areal mixture of the comet dark terrain (ROI 1) and 17\% of water ice (grain size of 30 $\mu$m).}
\label{scarpcolor}
\end{figure*}

Interestingly, we see important morphological changes of the surface near the Anhur/Bes boundary. A new 10-m high scarp has been formed near the original site of the Bes bright spot (patch B), as shown in Fig.~\ref{scarp} where  we compare images taken from May 2015 to 30 January 2016. We are not sure about the precise time of the formation of the scarp because the Rosetta spacecraft was moved into larger orbits from June 2015 to the end of 2015 due to perihelion passage with the consequent increased cometary activity. A careful inspection of the images in the OSIRIS archive indicates that the scarp was not yet formed on 1 August 2015, but already present on 10 December 2015 images (Fig.~\ref{scarp}). From the images acquired at higher resolution on January 2016, we evaluate the length of this scarp to be about 140 m. It bounds a depressed area with an extension of 4000-5000 m$^2$, suggesting a maximum volume loss of about 50000 m$^3$ from sublimation. The collapse event most likely occurred due to the comet's enhanced activity close to the perihelion. \\
At the foot of the scarp and within the newly formed depression, the deposits display more boulders compared 
to the original terrains at the scarp hanging wall. Thus this event generated a  collapse of the material with the formation of new boulders.

The new scarp was observed on 7 May 2016 from a spacecraft distance of 11 km, corresponding to a spatial resolution of 0.2 m/px, with 7 NAC filters. It shows bluer colours than the surrounding area (Fig.~\ref{scarpcolor}). Notably the freshly exposed terrain indicated by the regions of interest (ROI) 2 and 3 (Fig.~\ref{scarpcolor}) is less red than ROI 1 located on the top of the scarp (ROI 1), while the bright region in ROI 4 (blue square symbol), located close to the shadow of a $\sim$ 9m $\times$ 6 m boulder, has a flat spectrophotometric behaviour indicating the presence of some water ice. Several isolated bright spots with a reflectance similar to that of ROI 4 are also visible close to the other boulders located in the new scarp.  We estimate the water ice content in ROI 4  using a simple areal mixture model. To do so, we first compute the normal albedo of the ROIs using the geometric information derived from the shape 7S model (Jorda et al., 2016), and the Hapke model parameters determined by Fornasier et al. (2015, their table 4) from resolved photometry in the orange filter centred at 649 nm. We assume that the phase function at 649 nm also applies at the other wavelengths. To simulate the observed reflectance of ROI 4, we create an areal mixture of two components: the comet dark terrain, represented by ROI 1 at the top of the new scarp, and water ice. The water ice spectrum was derived from the synthetic reflectance from Hapke modelling starting from optical constants published in Warren and Brandt (2008) and adopting a grain size of 30 $\mu$m, that is a typical size for ice grains on cometary nuclei (Sunshine et al., 2006; Capaccioni et al., 2015; Filacchione et al., 2016a). This assumption was necessary because it is not possible to constrain the ice particle size from OSIRIS observations. The model that best fits the reflectance of ROI 4 is the areal mixture of the comet dark terrain enriched with 17$\pm$2\% of water ice, i.e. a value comparable to what has been found for the patch B seen one year earlier in the Bes region.

\subsection{Spectral slope evolution}

\begin{figure*}
\centering
\includegraphics[width=1.0\textwidth]{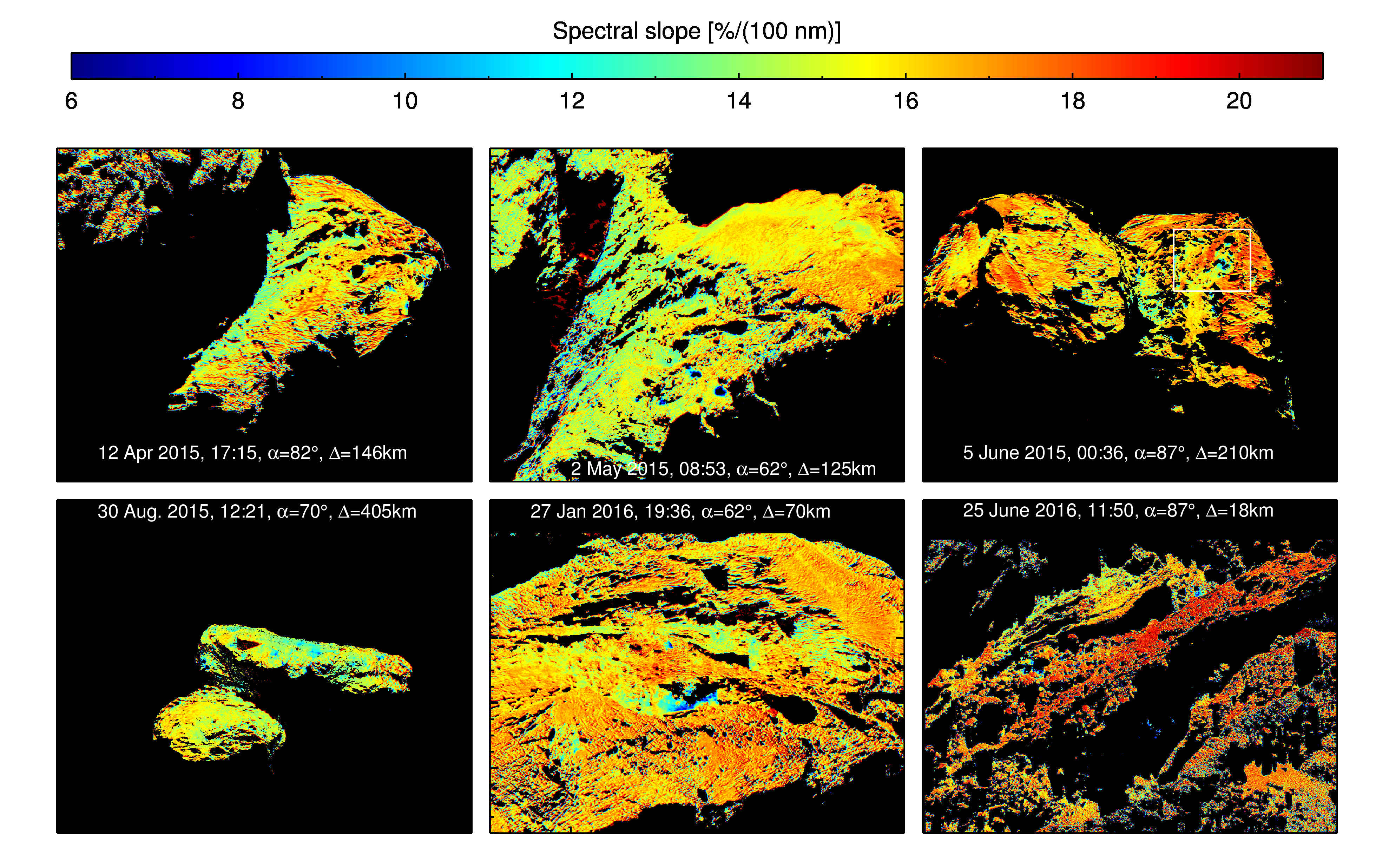}
\caption{Spectral slope, evaluated in the 535-882 nm range, from different images in the April 2015-June 2016 time frame. The white rectangle on June 2015 image indicates the region seen with a spatial resolution 10 times higher on June 2016 images acquired at the same phase angle.}
\label{slopes}
\end{figure*}

The evolution of the spectral slope of the Anhur region is shown in Fig.~\ref{slopes}. As the observations are obtained at different phase angles and the phase reddening has been found to be important for the 67P's nucleus and variable with the cometary activity (Fornasier et al., 2015, 2016; Feller et al., 2016), the analysis of the spectral evolution over time is not obvious. Several areas having a lower spectral slope value are evident in Fig.~\ref{slopes} (relatively bluer surfaces in the given colour scale) and associated with regions enriched in water ice and/or active. The comparison of images acquired at very similar phase angles clearly shows an increase of the spectral slope for images acquired post-perihelion at larger heliocentric distances. The 2 May 2015 and 27 January 2016 data were both acquired at phase angle 62$^{\circ}$, and the average spectral slope in the 535-882 nm range increased from $S$=14.9 \%/(100 nm) to 16.4 \%/(100 nm) from 2015 to 2016 observations. Local spectral slope variations (notably regions with a much lower spectral slope value) are associated with the large water-ice-rich patches (on 2 May 2015 images) or to the exposed ice close to a scarp previously discussed (on 27 January 2016). Similarly, the data acquired at phase angle 87$^{\circ}$ show the same trend, with the spectral slope $S$ increasing from 15.9 \%/(100 nm) to 17.5 \%/(100 nm) from June 2015 to June 2016 data in the observed common region. Fornasier et al. (2016) analysed the spectral slope evolution over time and heliocentric distance for the comet finding a progressive spectral slope decrease of the entire comet surface as it approached the Sun when the intense activity lifted off large amounts of dust, partially exposing the ice-rich terrain underneath, followed by a spectral slope increase in the post-perihelion orbit at heliocentric distance $>2$ au. The spectral slope evolution observed for the Anhur-Bes regions confirms the results reported by Fornasier et al. (2016) for the whole comet, notably the continuous changes of the physical properties of the uppermost layers of the nucleus related to the cometary heliocentric distance and to the level of activity. As observed for the whole comet, the Anhur-Bes regions also became spectrally redder at large post-perihelion distances when the activity was no longer capable of sustained removal of dust. Keller et al. (2017) estimates that up to 20\% of the dust particles lifted up by the cometary activity from the active Southern hemisphere during the perihelion passage may directly return to the nucleus surface with re-deposition in the Northern hemisphere, implying important mass transfer between the 2 hemispheres. However even if the southern hemisphere is depleted in large scale smooth terrains and is dominated by consolidated material (El Maarry et al., 2016), a dust mantle covers its surface with a thickness varying with the cometary activity, thus explaining the red colours observed in the Bes and Anhur regions beyond $\sim$ 2 au.

\section{Conclusions}

In this work, we have investigated the geomorphology, the spectrophotometric properties, and the transient events of Anhur and part of Bes regions. They are most fragmented regions showing fractured and patterned rough terrains, and displaying talus deposits, elongated canyon-like structures, scarp retreats, diamictons, and degradated sequences of strata indicating a pervasive layering. They receive an intensive insolation during the the Southern summer. Because of this, the erosion rate due to the sublimation of water ice in the Southern hemisphere is up to 3-4 times higher than that in the Northern hemisphere (Keller et al., 2015, 2017). Thus Southern regions give potentially a better glimpse of the pristine cometary material than the Northern ones, where the dust coating is thicker.\\
 The Anhur/Bes regions are highly active and are sources of several jets, including the strongest outburst event ever observed, as established in this paper and in Vincent et al. (2016a). Most of the jets's sources are found close to the Anhur/Aker boundary, i.e. close to cliffs. Anhur also hosts the only active pit observed in the Southern hemisphere. 

The Anhur/Bes regions are sculpted by staircase terraces that support the nucleus stratification hypothesis formulated by Massironi et al. (2015). Anhur seems highly eroded with the presence of several scarps dissecting the different strata. Interestingly, at the feet of scarps and cliffs, we observed both talus and gravitational accumulation deposits. These deposits often show a relatively bluer spectral behaviour than the surroundings, pointing to an enrichment in the surface water ice content, and in some cases a higher reflectance and a flat spectrophotometric behaviour, consistent with the presence of exposed water ice. These deposits are sometimes also sources of activity. These observations thus reinforce the hypothesis by Vincent et al. (2016b) that the fresh material  falling from cliffs/scarps is volatile rich and may become active.

In the boundary between Anhur and Bes, two water-ice-rich patches were visible for about 10 days (between at least 27 April and 7 May 2015), shortly before the equinox, and they were observed one month after the unique detection of exposed CO$_2$ ice on the 67P's nucleus (Filacchione et al., 2016), in the location corresponding to one of the two bright patches (patch A). These ice-rich patches formed in a smooth terrace covered by a layer of fine deposits on the consolidated material. The fact that first the CO$_2$ ice and then the H$_2$O ice was exposed, indicates a progressive stratification of different volatiles resulting from recondensation and sintering of the subsurface material during previous perihelion passages, and clearly points to local compositional heterogeneities on scales of several tens of meters. \\
In this peculiar Anhur/Bes boundary we also noticed a new scarp formed sometime between the perihelion passage and December 2015. The scarp is about 140 m long and 10 m high, bounding a depressed area of about 4000-5000 m$^2$, and generated a collapse of the material with the formation of new boulders. The strong activity throughout the perihelion passage, together with the observed local surface and subsurface enhancement in volatiles in these areas presumably triggered the formation of this new scarp. Also in this case, the freshly exposed material collapsed from the scarp shows a bluer colour and a lower spectral slope indicating the presence of some water ice, reaching abundance of about 17\% in the shadows of some boulders located in the new depression. 

The spectral slope evolution from April 2015 to June 2016 clearly indicates that the Anhur/Bes regions, as observed for other regions of the comet, became spectrally redder post-perihelion at heliocentric distances $>$ 2.0 au compared to the pre-perihelion data. This indicates continuous changes of the physical properties of the uppermost layers, in terms of composition and roughness, driven by the different levels of cometary activity and insolation. Close to perihelion the strong cometary activity thinned out the nucleus dust, partially exposing the underlying ice-rich layer, resulting in lower spectral slope values seen all over the nucleus as shown by Fornasier et al. (2016). This implies that water ice is abundant just beneath the surface on the whole nucleus, as also inferred by the study of the gaseous emission in the coma, with the H$_2$O emission which is strongly correlated with the local illumination on the nucleus, and produced predominantly by local well-illuminated active areas (Bockelee-Morvan et al., 2015; Migliorini et al., 2016). Far from the Sun, as the cometary activity decreased, the dust mantle became thicker, masking the subsurface ice-rich layer. This results in a steep spectral slope associated with a surface composition depleted in volatiles.

\vspace{0.3truecm}
{\bf Acknowledgments} \\

OSIRIS was built by a consortium led by the Max-Planck-Institut f\"ur Sonnensystemforschung,Goettingen, Germany, in collaboration with CISAS, University of Padova, Italy, the Laboratoire d'Astrophysique de Marseille, France, the Instituto de Astrof\'isica de Andalucia, CSIC, Granada, Spain, the Scientific Support Office of the European Space Agency, Noordwijk, The Netherlands, the Instituto Nacional de T\'ecnica Aeroespacial, Madrid, Spain, the Universidad Polit\'echnica de Madrid, Spain, the Department of Physics and Astronomy of Uppsala University, Sweden, and the Institut  f\"ur Datentechnik und Kommunikationsnetze der Technischen Universitat  Braunschweig, Germany.
The support of the national funding agencies of Germany (DLR), France (CNES), Italy (ASI), Spain (MEC), Sweden (SNSB), and the ESA Technical Directorate is gratefully acknowledged. 
We thank the Rosetta Science Ground Segment at ESAC, the Rosetta Mission Operations Centre at ESOC and the Rosetta Project at ESTEC for their outstanding work enabling the science return of the Rosetta Mission. The authors thanks Dr. E. Howell for her comments and suggestions which helped us to improve this article.
\bigskip





\section*{Affiliations}
$^1$ LESIA, Observatoire de Paris, PSL Research University, CNRS, Univ. Paris Diderot, Sorbonne Paris Cit\'{e}, UPMC Univ. Paris 06, Sorbonne Universit\'es, 5 Place J. Janssen, 92195 Meudon Principal Cedex, France\\
$^2$ Department of Earth Sciences, National Central University, Chung-Li 32054, Taiwan\\
$^3$ Dipartimento di Geoscienze, University of Padova, via G. Gradenigo 6, 35131 Padova, Italy\\
$^4$ Center of Studies and Activities for Space (CISAS) {\it G. Colombo}, University of Padova, Via Venezia 15, 35131 Padova, Italy\\
$^5$ Laboratory for Atmospheric and Space Physics, University of Colorado, 3665 Discovery Drive, CO 80301, USA\\
$^6$ Deutsches Zentrum f\"ur Luft und Raumfahrt (DLR), Institut f\"ur Planetenforschung, Asteroiden und Kometen, Rutherfordstrasse 2, 12489 Berlin, Germany\\
$^7$ Institut f\"ur Geophysik und extraterrestrische Physik (IGEP), Technische Universitat Braunschweig, Mendelssohnstr. 3, 38106 Braunschweig, Germany\\
$^8$ National Central University, Graduate Institute of Astronomy, 300 Chung-Da Rd, Chung-Li 32054 Taiwan\\
$^9$ Space Science Institute, Macau University of Science and Technology, Macau, China\\
$^{10}$ Max-Planck-Institut f\"ur Sonnensystemforschung, Justus-von-Liebig-Weg, 3, 37077, G\"ottingen, Germany\\
$^{11}$ University of Padova, Department of Physics and Astronomy, Vicolo dell'Osservatorio 3, 35122 Padova, Italy\\
$^{12}$ Laboratoire d'Astrophysique de Marseille, UMR 7326 CNRS, Universit\'e Aix-Marseille, 38 rue Fr\'ed\'eric Joliot-Curie, 13388 Marseille Cedex 13, France\\ 
$^{13}$ Centro de Astrobiologia, CSIC-INTA, 28850, Torrejon de Ardoz, Madrid, Spain\\
$^{14}$ International Space Science Institute, Hallerstrasse 6, 3012 Bern, Switzerland\\
$^{15}$ Scientific Support Office, European Space Research and Technology Centre/ESA, Keplerlaan 1, Postbus 299, 2201 AZ Noordwijk ZH, The Netherlands\\
$^{16}$ Department of Physics and Astronomy, Uppsala University, Box 516, 75120 Uppsala, Sweden\\
$^{17}$ PAS Space Reserch Center, Bartycka 18A, 00716 Warszawa, Poland\\
$^{18}$ University of Maryland, Department of Astronomy, College Park, MD 20742-2421, USA\\
$^{19}$ LATMOS, CNRS/UVSQ/IPSL, 11 boulevard d'Alembert, 78280, Guyancourt, France\\
$^{20}$ INAF, Osservatorio Astronomico di Padova, Vicolo dell'Osservatorio 5, 35122 Padova, Italy\\
$^{21}$ CNR-IFN UOS Padova LUXOR, Via Trasea, 7, 35131 Padova , Italy\\
$^{22}$ NASA Jet Propulsion Laboratory, 4800 Oak Grove Drive, Pasadena, CA 91109, USA\\
$^{23}$ Department of Mechanical Engineering, University of Padova, via Venezia 1, 35131 Padova, Italy\\
$^{24}$ University of Trento, Via Mesiano 77, 38100 Trento, Italy\\
$^{25}$ INAF-Osservatorio Astronomico, Via Tiepolo 11, 34014 Trieste, Italy\\
$^{26}$ Aix Marseille Universit\'e, CNRS, LAM (Laboratoire d'Astrophysique de Marseille), UMR 7326, 38 rue Fr\'ed\'eric Joliot-Curie, 13388 Marseille Cedex 13, France\\
$^{27}$ Instituto  de Astrofisica de Andalucia (CSIC), c/ Glorieta de la Astronomia s/n, 18008 Granada, Spain\\
$^{28}$ Operations Department, European Space Astronomy Centre/ESA, P.O.Box 78, 28691 Villanueva de la Canada, Madrid, Spain\\
$^{29}$ University of Padova, Department of Information Engineering, Via Gradenigo 6/B, 35131 Padova, Italy\\
$^{30}$ NASA Ames Research Center, Moffett Field, CA 94035, USA\\
$^{31}$ Physikalisches Institut der Universit\"at Bern, Sidlerstr. 5, 3012 Bern, Switzerland\\
$^{32}$ MTA CSFK Konkoly Observatory, Budapest, Hungary

\bsp    
\label{lastpage}
\end{document}